\documentclass{aa}  

\usepackage{graphicx}
\usepackage{float}
\usepackage{wrapfig}
\usepackage{ragged2e}
\usepackage{geometry}
\usepackage{txfonts}
\usepackage{subcaption}
\usepackage{hyperref}
\usepackage{ulem}
\usepackage[dvipsnames]{xcolor}
\hypersetup{
    colorlinks = true,
    citecolor ={cyan}
}
\def\rsun{{~R}_{\odot}}
\def\msun{{~M}_{\odot}}
\def\zsun{{~Z}_{\odot}}

\newcommand{\days}{\ensuremath{\rm \,days}}

\newcommand{\msunyr}{\ensuremath{\rm \,M_\odot \, yr^{-1}}}
\newcommand{\logmdot}{\ensuremath{{\rm log}(\dot{M})}}
\newcommand{\mdot}{\ensuremath{\dot{M}}}

\bibpunct[; ]{(}{)}{;}{a}{}{;}

\definecolor{darkgreen}{rgb}{0.0, 0.4, 0.22}

\newcommand{\logteff}{\ensuremath{\,{\rm log} (T_{\rm eff} / {\rm K})}}
\newcommand{\logg}{\ensuremath{\,{\rm log} (g)}}
\newcommand{\logL}{\ensuremath{\,{\rm log} (L / L_{\odot})}}

\usepackage{array}
\newcolumntype{L}[1]{>{\raggedright\let\newline\\\arraybackslash\hspace{0pt}}m{#1}}
\newcolumntype{C}[1]{>{\centering\let\newline\\\arraybackslash\hspace{0pt}}m{#1}}
\newcolumntype{R}[1]{>{\raggedleft\let\newline\\\arraybackslash\hspace{0pt}}m{#1}}

\begin{document} 

   \title{Partial-envelope stripping and nuclear-timescale mass transfer from evolved supergiants at low metallicity}
   \titlerunning{Nuclear-timescale mass transfer from evolved massive stars}
   \authorrunning{Klencki et al.}
   
      \author{Jakub Klencki\inst{1,2}
          \and
          Alina Istrate\inst{2}
          \and
          Gijs Nelemans\inst{2,3,4}      
          \and 
          Onno Pols\inst{2}
   }

   \institute{
   European Southern Observatory, Karl-Schwarzschild-Strasse 2, 85748 Garching bei München, Germany\\
     \email{jakub.klencki@eso.org}
     \and
   Department of Astrophysics/IMAPP, Radboud University, P O Box 9010, NL-6500 GL Nijmegen, The Netherlands
         \and
          Institute of Astronomy, KU Leuven, Celestijnenlaan 200D, B-3001 Leuven, Belgium
         \and
         SRON, Netherlands Institute for Space Research, Sorbonnelaan 2, NL-3584 CA Utrecht, The Netherlands
   }
   \date{Received November, 2021; accepted...}

  \abstract
   {Stable mass transfer from a massive post-main sequence (post-MS) donor is thought to be a short-lived event of thermal-timescale mass transfer ($\sim 10^{-3}\msunyr$) which within $\lesssim 10^4 \, \rm yr$ strips the donor star of nearly its entire H-rich envelope, producing a hot, compact helium star. 
   This long-standing picture is based on stellar models with rapidly-expanding Hertzprung gap (HG) donor stars. 
   Motivated by a finding that in low-metallicity binaries, post-MS mass transfer may instead be initiated by donors already at the core-helium burning (CHeB) stage, we use the MESA stellar-evolution code to compute grids of detailed massive binary models at three metallicities: those of the Sun, the Large Magellanic Cloud (LMC, $Z_{\rm Fe;LMC} / Z_{\rm Fe; \odot} \approx 0.36$), and the Small Magellanic Cloud (SMC, $Z_{\rm Fe;SMC} / Z_{\rm Fe; \odot} \approx 0.2$). Our grids span a wide range in orbital periods ($\sim 3$ to $5000$ days) and initial primary masses ($10 \msun$ to $36$-$53 \msun$, depending on metallicity).
   
   We find that metallicity strongly influences the course and outcome of mass-transfer evolution. We identify two novel types of post-MS mass transfer: (a) mass exchange on the long nuclear timescale  ($\Delta T_{\rm MT} \gtrsim 10^5 \, \rm yr$, $\dot{M} \sim 10^{-5}\msunyr$) that continues until the end of the CHeB phase, and (b) rapid mass transfer leading to detached binaries with mass-losers that are only partially stripped of their envelopes.  At LMC and SMC compositions, the majority of binary models with donor masses $\geq 17\msun$ follow one of these two types of evolution.
   In neither (a) or (b) does the donor become a fully stripped helium star by the end of CHeB. Boundaries between the different types of post-MS mass transfer evolution are associated with the degree of rapid post-MS expansion of massive stars and, for a given metallicity, are sensitive to the assumptions about internal mixing. 
   
   At low metallicity, due to partial envelope stripping, we predict fewer hot fully stripped stars formed through binary interactions as well as higher compactness of the pre-supernova core structures of mass losers. Nuclear-timescale post-MS mass transfer suggests a strong preference for metal-poor host galaxies of ultra-luminous X-ray sources with black-hole (BH) accretors and massive donors, some of which might be the immediate progenitors of binary BH mergers.  
   It also implies a population of interacting binaries with blue and yellow supergiant donors. 
   Partially-stripped stars could potentially explain the puzzling nitrogen-enriched slowly-rotating (super)giants in the LMC. 
   }
  
    \keywords{stars: massive -- stars: binaries: general -- stars: evolution}
   \maketitle

\section{Introduction}
\label{sec:intro}

Massive stars do not live alone. The majority are formed in close binary or higher-order systems
in which they are destined to strongly interact with their companions by transferring mass and
angular momentum \citep{Sana2012,Moe2017}. 
Mass transfer has significant consequences for basic properties and final fates of both stars.
Many mass losers become helium-rich stripped stars,
from hot subdwarfs to Wolf-Rayet (WR) stars,  \citep[depending on the mass;][]{Paczynski1967,vdHeuvel1975,Vanbeveren1991,Podsiadlowski1992,deLoore1992,Petrovic2005,Eldridge2008,Gotberg2017,Laplace2020},
and a UV-bright prominent source of ionizing photons 
\citep{Gotberg2018,Gotberg2020_ionizing}. 
Mass gainers and stellar mergers, on the other hand, are the likely explanation 
for the existence of blue stragglers \citep{McCrea1964,Pols1994, Braun1995,Glebbeek2013,Schneider2015}, may dominate the population 
of rapidly rotating stars \citep{deMink2013,Renzo2021} in particular Be stars \citep{Pols1991,Shao2014,Hastings2021}, lead to 
formation of magnetic fields in stars \citep{Ferrario2009,Schneider2019Natur}, 
and a peculiar class of supernovae from blue-supergiant progenitors \citep[including the most recent naked-eye supernova SN 1987A, see][]{Podsiadlowski1990,Justham2014}.
Prior mass transfer interactions have been shown to leave clear signatures 
in the properties of stellar cores at core collapse, affect lightcurves and yields of the resulting supernovae, 
and play an important role in deciding whether the final remnant will be a
neutron star (NS) or a black hole \citep[BH, ][]{Morris2007,Justham2014,Woosley2019,Schneider2021,Laplace2021,Vartanyan2021}.
Through mass transfer, binaries enrich the interstellar medium with processed material \citep{deMink2009},
shine as X-ray binaries \citep{vdHeuvel1973,vdHeuvel1975,Verbunt1993},
and bring two BHs or neutron stars NSs close together, driving them to merge and blaze in gravitational waves \citep[GW;][]{Tutukov1993,PortegiesZwart1998,Belczynski2002,Voss2003} 
in the most energetic spectacles observed to date
\citep{Abbott16_firstBBH}.

In most 
massive binary systems, 
a phase of mass transfer is initiated when one of the stars evolves 
off the main sequence (MS) and expands significantly leading to the so-called 
case B Roche-lobe overflow \citep[RLOF, ][]{Kippenhahn1967,Paczynski1971}.
\footnote{There is some ambiguity in the literature as to whether the term case B mass transfer refers to only 
the cases initiated before the onset of core-He burning or also during that phase of evolution. Throughout 
this paper, we will adapt the latter naming convention.}
The textbook view of case B mass transfer is that the donor star looses nearly its entire H-rich envelope 
in a short-lived phase of thermal-timescale mass exchange \citep[$\lesssim 10^4$ yr][]{Kippenhahn1967,Paczynski1971,vdHeuvel1975,Podsiadlowski1992,Vanbeveren1998}.
This is because the donor is thought to usually be
a rapidly-expanding H-shell burning giant, a Hertzprung gap (HG) star. It is out of 
thermal-equilibrium even before any RLOF occurs and would need to expand all the way to the red giant 
branch in order to regain a stable thermal structure. In a close binary, this continuous rapid expansion of the donor star
is what causes the mass transfer to proceed on the thermal timescale until only a thin envelope layer remains. 
By the time the binary detaches, the donor becomes a stripped star, composed predominantly of helium.

This paradigm may no longer be true for massive binaries in low-metallicity (low-Z) environments. As pointed out by \citet{deMink2008}
and recently shown in much detail by \citet{Klencki2020}, metallicity has a strong effect on what is the 
typical evolutionary state of donors when they initiate a case B mass transfer phase in a population of massive binaries. 
While at high (Solar-like) metallicity a post-MS interaction nearly always occurs when the donor is expanding as a HG star, it has been shown that
at low metallicity the parameter space for RLOF from core-He burning donors becomes very significant 
\citep[and possibly dominant above a certain mass, see][]{Klencki2020}. In contrast to HG donors, stars that are already 
at the core-He burning stage are in thermal equilibrium and slowly expanding on the nuclear timescale of $\sim 10^6$ yr. 
Mass transfer from such donors has never been studied with detailed binary evolution models and its understanding is still lacking.

At the same time, various observational clues suggest that metallicity has a strong 
influence on the evolution of massive stars and binaries. 
Long gamma-ray bursts \citep{Graham2013}, superluminous supernovae \citep{GalYam2012}, and ultra-lumionous X-ray (ULX) sources \citep{Kovlakas2020}
all preferentially occur in low-Z galaxies. 
Similarly, Type Ic supernovae with broad lines in the spectra 
are typically found in metal-poor hosts, whereas normal Type Ic supernovae avoid dwarf galaxies \citep{Modjaz2011}.
It has been suggested that BHs with masses above $\sim30\msun$, frequently found in binary BH mergers detected in GW by LIGO/Virgo \citep{gwtc2_2020}, may originate from metal-poor environments, where the reduced strength of line-driven winds allows for the formation of more massive BHs \citep[e.g.][]{Belczynski2010_maxBHmass,Vink2021}.
Thanks to their close proximity to the Milky Way, the Small and the Large Magellanic Cloud (SMC and LMC, respectively) serve as excellent test-beds of massive star evolution at low metallicity. Large-scale spectroscopic surveys of the Magellanic Clouds have revealed populations of stars that cannot be explained by current models, in particular the slowly-rotating nitrogen-enriched (super) giants \citep{Evans2006,Hunter2008,McEvoy2015,Grin2017}. The population of high-mass X-ray binaries (HMXBs) also appears to be metallicity-dependent, with a surprisingly large number of Be HMXBs found in the Magellanic Clouds \citep{Dray2006}.

Motivated by the importance of low metallicity on the evolution of massive stars, evidenced on both theoretical and observational grounds, in this work we follow up on \citet{Klencki2019,Klencki2020} and explore mass-transfer evolution in massive binaries of different metallicities.

The paper is organized as follows. In Section~\ref{sec.method} we describe our computational method and physical 
assumptions as well as the parameter space explored with our models. In Section~\ref{sec:results} we present the results from our binary evolution sequences, focusing on the mass-transfer evolution from post-MS donors. In Section~\ref{sec:why_partstrip} we take an in-depth look at the origin of different types of mass-transfer evolution found in Sec.~\ref{sec:results}. In Section~\ref{sec:discussion} we discuss the caveats of our models as well as various implications of the findings. We conclude in Section~\ref{sec:summary}. 

\section{Method: binary stellar evolution models}
\label{sec.method}

\subsection{Physical ingredients: single and binary evolution}
\label{sec.method_single_binary}

We employ the MESA stellar evolution code \citep{Paxton2011,Paxton2013,Paxton2015,Paxton2018,Paxton2019}
\footnote{MESA version r11554, \url{http://mesa.sourceforge.net/}}. Convection 
is modeled using the mixing-length theory \citep{BohmVitense1958} with the mixing length of $\alpha = 1.5$.
We employ the Ledoux criterion for convection and account for semiconvective mixing with an efficiency 
of $\alpha_{\rm SC} = 33$. We account for convective core-overshooting by applying step overshooting 
with an overshooting length $\sigma_{\rm ov} = 0.33$ \citep{Brott2011}. Such choices of overshooting 
and semiconvection efficiency were shown to be in good agreement with the observed ratio of
blue and red supergiants in the SMC \citep{Schootemeijer2019, Klencki2020}.

Models are computed at three different initial chemical compositions: either at Solar metallicity with 
$Z = 0.017$ and abundance ratios from \citet{Grevesse1996} or with non-Solar abundance ratios of the Magellanic 
Clouds following \citet{Brott2011}. In the case of the LMC that yields $Z_{\rm LMC} \approx 0.0048$ and the relative iron abundance 
$Z_{\rm Fe;LMC} / Z_{\rm Fe; \odot} \approx 0.36$, whereas in the case of SMC: $Z_{\rm SMC} \approx 0.0022$ and the 
relative iron abundance $Z_{\rm Fe;SMC} / Z_{\rm Fe; \odot} \approx 0.2$. 
Similarly to \citet{Langer2020}, we use custom-made OPAL opacity tables \citep{Iglesias1996} corresponding 
to the adopted initial abundances of the SMC or the LMC. 

The winds of stars on the cool side of the bi-stability jump 
with $T_{\rm eff} < 0.95~T_{\rm eff;jump} \approx 25$ kK (see Eqn.~15 in \citealt{Vink2001}) are modeled as 
the larger of the mass-loss rates from \citet{Vink2001} and \citet{Nieuwenhuijzen1990}. 
The winds of stars on the hot side of the bi-stability jump (with $T_{\rm eff} > 1.05~T_{\rm eff;jump}$)
are modeled as a combination 
of several different prescriptions, depending on the surface hydrogen abundance $X$.
For stars with $X > 0.45$, we follow \citet{Vink2001}. For stars with $ 0.1 < X < 0.35$, we apply 
the empirical mass-loss rates from \citet{Nugis2000}. 
For (nearly) hydrogen-free stars with $X < 0.05$, we follow \citet{Yoon2017}, whose prescriptions for WNE 
stars are based on the results from \citet{Hainich2014} and for WC/WO stars were derived by \citet{Tramper2016}.
In the intermediate $X$ regimes, as well as in the temperature range around the bi-stability jump 
$0.95 < T_{\rm eff}/T_{\rm eff;jump} < 1.05$, we linearly interpolate between the above prescriptions to provide smooth transitions. 

\begin{figure*}
    \includegraphics[width=\textwidth]{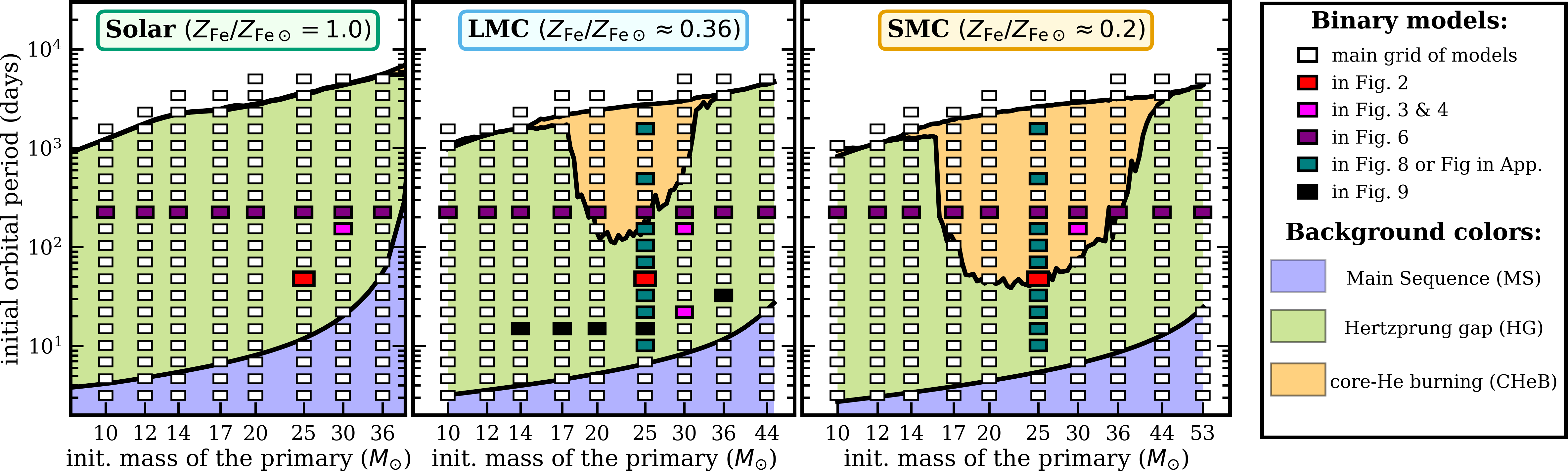}
\centering
\caption{Distribution of the main grid of binary models at each metallicity over the parameter space of
initial orbital periods and primary masses. Background color indicates the evolutionary 
stage of the primary star at the onset of mass transfer through RLOF. We differentiate between 
core-H burning donors (blue), H-shell burning donors (i.e.
donors that experience the Hertzprung-gap phase of rapid expansion; green), as well as core-He burning donors (yellow). 
The initial mass ratio is $q = M_2 / M_1 = 0.6$. 
Colored rectangles indicate which models from the grid are shown in figures throughout the paper (see legend). 
}
\label{fig.bin_param_ranges}
\end{figure*}

In addition, when using the \citet{Vink2001} prescription at $T_{\rm eff} > T_{\rm eff;jump}$ (optically thin winds of OB supergiants), 
we account for the possible transition to the optically thick WR-type mass loss. 
Both theoretical \citep{Grafener2008,Vink2011,Sander2020a} and empirical studies \citep{Grafener2011,Bestenlehner2014}
suggest that this transition takes place when the Eddington factor 
$\Gamma_{\rm e}=10^{-4.813}(1+X)(L/L_{\odot}) (M/M_{\odot})^{-1}$ becomes sufficiently large, 
which typically happens when a substantial amount of mass has already been lost (either in winds 
or as a result of mass transfer). 
Here, we follow \citet{Grafener2008} and assume that this threshold is at $\Gamma_{\rm e} \approx 
0.1 + \Gamma_{\rm 0}$ where $\Gamma_{\rm 0}$ is metallicity (iron) dependent 
and equals  to $0.326$, $0.468$, and $0.514$ for our Solar, LMC, and SMC iron abundances, respectively
(Eqn.~6 of \citealt{Grafener2008}). 
We maintain the \citet{Vink2001} mass-loss rate for $\Gamma_{\rm e} - \Gamma_{\rm 0} < 0.08$, 
whereas for $\Gamma_{\rm e} - \Gamma_{\rm 0} > 0.12$ we apply the theoretical recipe for optically thick 
winds from \citet{Grafener2008}. For intermediate Eddington factors, we linearly interpolate between the two mass-loss rates. 

We include rotationally-induced mixing of elements due to Eddington-Sweet circulation, secular 
shear instabilities, and the Goldreich-Schubert-Fricke instability, with an efficiency factor $f_c = 1/30$ \citep{Heger2000,Brott2011}.
We model rotationally-enhanced mass loss as in \citet{Langer1998}, see implementation in \citet{Paxton2013}.
In most of our models, 
we avoid using the MLT++ option in MESA \citep{Paxton2013} as it was shown to artificially 
reduce the stellar radii during 
the giant phases of evolution and therefore could affect the behavior of stars during mass transfer. 
In the few cases of the most massive donors in our grid ($36 \msun$, $44 \msun$, and $53 \msun$ at Solar, LMC, and SMC compositions, 
respectively) we resort to a limited use of MLT++ by gradually 
reducing superadiabacity in outer layers ($T < 10^6$ K) of donors once they have reached the mass-transfer rate of $\logmdot = -2.5$. 
These models would have not converged otherwise due to numerical difficulties arising at the bottom of a subsurface convective zone 
located at the iron opacity peak.

We follow the formalism of \citet{Kolb1990} to calculate the mass-transfer rate through the L1 
Lagrangian point.
This includes both the optically-thin regime of mass transfer when $R_{\rm don}$ is still slightly smaller than $R_{\rm RL}$, 
where $R_{\rm don}$ is the radius of the donor star and $R_{\rm RL}$ is the size of its Roche lobe, as well as the 
main phase of mass transfer when $R_{\rm don} > R_{\rm RL}$.
The mass transfer is assumed to remain stable for as long as the binary model computes, i.e. unstable cases are when the 
mass transfer rate clearly diverges to infinity and the simulation stops. 

The main focus of the paper is to study the behavior of the donor star during a phase of mass transfer 
and to understand how the slowed down post-MS expansion of low-metallicity massive stars influences their
evolution as donors in interacting binaries. 
Therefore, for simplicity, in our binary models we only evolve the primary (donor) star and 
treat the companion as a point mass. 
We also assume that the accretion efficiency is Eddington-limited (calculated for a BH accretor), which in practice will mean nearly fully non-conservative mass transfer. We further assume that the specific 
angular momentum of the mass ejected from the system is that of the accretor on its binary orbit (i.e. the isotropic re-emission mode). 
These assumptions make our models well suited to represent systems with compact-object accretors (specifically BHs, given that our accretors are always more massive than $6\msun$). However, as we argue in Sec.~\ref{sec.applicability_stellar_binaries}, all our main findings and conclusions also hold in the case of systems with stellar accretors. Throughout the paper we will thus not assume anything about the nature of the companion unless explicitly stated otherwise.

We model the spin-up of the primary due to tidal interactions following the synchronization timescale for radiative envelopes
from \citet{Hurley2002}. In MESA the angular momentum from tides is being deposited in such a way that 
each layer is being synchronized on the same timescale. We include the effect of isotropic wind mass-loss 
on the evolution of 
orbital parameters as well as the spin of the mass-losing star. The computed binary models as well as MESA inlists (input files) used in this work are available at \url{https://zenodo.org/record/6412508}.

\subsection{Initial parameter space and stopping conditions}
\label{sec.method_param_space}

\begin{figure*}[h]
\includegraphics[width=\textwidth]{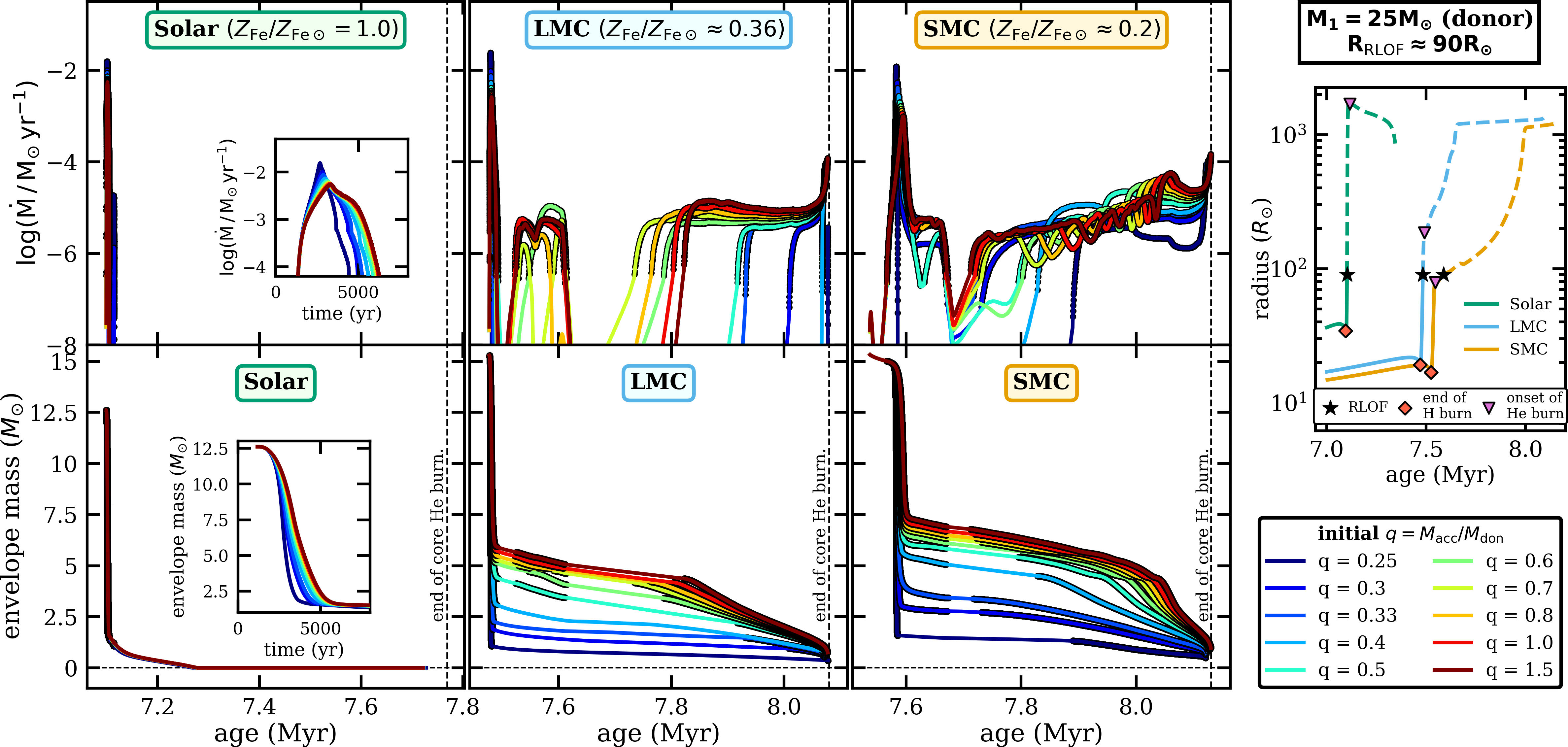}
\caption{
Mass transfer evolution in binaries with a $25 \msun$ primary (initial mass), compared between the three different metallicities. 
Top panels show the mass-transfer rate, bottom panels show the mass of the envelope
(hydrogen content $X_H > 10^{-3}$).
Different colors correspond to different mass ratios 
between the binary components $q = M_2/M_1$, varying from $0.1$ and $1.5$ (the donor stays the same). Bold part of each curve indicates when the donor overflows its Roche lobe.
In each case, the initial RLOF takes place when the primary radius is around $R_{\rm RLOF} \approx 90 \rsun$, 
see the radius-age diagram on the right-hand side. 
In models with $q \leq 0.225$ (not shown in the figure) the mass transfer became dynamically unstable. 
Models terminate at the end of core-He burning.}
\label{fig.diffQ}
\end{figure*}

We compute three main grids of binary models: at Solar, LMC, and SMC composition. 
Each grid spans 20 different initial orbital periods $P_{\rm ini}$ ranging 
from $\approx 3.16$ to $5012$ days, spaced in equal logarithmic steps. 
The SMC grid spans 10 different initial primary masses $M_{\rm 1}$: $10$, $12$, $14$, $17$, $20$, 
$25$, $30$, $36$, $44$, and $53 \msun$. The LMC grid ends at $44 \msun$ whereas at Solar metallicity 
we compute models until $36 \msun$. 
The default initial mass ratio $q = M_{\rm 2} / M_{\rm 1}$ is $q = 0.6$ for our main grids. 
For the case of $M_1 = 25 \msun$ and $P_{\rm ini} \approx 47$ days we compute additional models 
with the mass ratio $q$ varying from $0.1$ to $1.5$ (15 models). We do this by preserving the 
structure of the donor at RLOF taken from the standard binary model with $q = 0.6$ and manually 
adjusting the mass ratio and orbital separation in such a way that at the moment of RLOF 
the donor has the same radius for any $q$. This approximation is justified because 
the tidal spin-up of the donor star, which would have been different in the evolution prior to RLOF 
for different mass ratios, is insignificant for binaries this wide. 
The primaries are initialized with a small initial rotational velocity of $30$ km/s. All the binaries 
are circular.

Figure~\ref{fig.bin_param_ranges} illustrates how our grids of binary models cover the parameter space for a mass transfer interaction 
initiated by donors at different evolutionary stages, similarly to figures in \citet{Klencki2020}.
At each composition, the grid covers the entire range of periods for a post-MS mass transfer interaction and fully encloses the range of masses at which donors 
begin the core-He burning stage as blue supergiants (at LMC and SMC compositions). 
In addition, each grid includes a small number of binaries that evolved through a case-A mass-transfer phase (i.e. initiated by a MS star).
These models are beyond the main scope of the paper and will not be discussed in detail. 

All the binary sequences terminate at central helium depletion ($Y_{\rm center} < 10^{-6}$). 
The remaining lifetimes of massive stars from that point on until the final core-collapse are short compared 
to hydrogen- and helium-burning lifetimes ($< 10^5$ yr or even $< 10^4$ yr for the more massive 
among our primaries).

\section{Results}
\label{sec:results}

\subsection{Nuclear-timescale mass transfer and partial envelope stripping}

We begin describing the results by taking a close look at binary evolution models from the middle of our grids:
systems with initial primary masses of $M_1 = 25\msun$ and orbital periods $P_{\rm ini} \approx 47\days$.
For this selected case of $M_1$ and $P_{\rm ini}$, apart from the default initial mass ratio $q = M_2 / M_1 = 0.6$, 
we computed additional mass transfer sequences with $q$ varying from $0.1$ to $1.5$, as described in 
Sec.~\ref{sec.method_param_space}. In each case, the size of the primary (donor) star at the moment of RLOF was about $R_{\rm RLOF} \approx 90\rsun$.
At both Solar and LMC compositions, the primary star of that radius was still a rapidly-expanding HG giant, whereas 
at SMC composition the donor was already at the core-He burning stage. 
For this selected set of models, in Fig.~\ref{fig.diffQ} we plot the time evolution of the mass transfer rate $\dot{M}$ (upper panels) 
and envelope masses (lower panels). Phases of RLOF are marked in bold. Other phases with $\logmdot > -8$ indicate mass transfer through stellar winds from a donor that is close to Roche-lobe filling. In all the SMC models, there is a small peak of wind mass transfer with $\logmdot \approx -7.2$ at the age of $7.55$ Myr, followed by a brief drop in $\dot{M}$ before RLOF starts at $7.6$ Myr. This is due to a temporary slight contraction of the primary after it regains equilibrium as a core-He burning star at $\sim 7.55$ Myr.

We find a remarkable difference in the mass-transfer evolution between the Solar case and the low-Z models of 
the LMC and SMC compositions. At Solar Z, irrespective of the mass ratio, the mass transfer interaction is a short-lived
and rapid event. Over the course of only several thousand years the donor star is stripped of its nearly entire H-rich 
envelope. This is a phase of thermal-timescale mass transfer with the rate reaching $\mdot \sim 10^{-2} \msunyr$. 
The thin remaining envelope layer ($\sim 1 \msun$) is further quickly lost in winds and the primary becomes a fully 
stripped helium star.
This type of evolution is the canonical picture of how a case-B mass transfer interaction can lead to the envelope loss 
and formation of stripped stars of various kinds, from subdwarfs to WR stars (see Sec.~\ref{sec:intro} and the references therein).

However, the evolution of LMC and SMC models in Fig.~\ref{fig.diffQ} turns out to be substantially different. 
In these low-Z cases the initial rapid (thermal) phase 
of mass transfer slows down or terminates while a significant part of the envelope is still retained ($M_{\rm env;left}$ from $\sim 1$ up to even $7 \msun$, 
depending on the mass ratio and metallicity). In models with $M_{\rm env;left} \gtrsim 2.5 \msun$, 
when the partially-stripped donor continues its evolution as a core He-burning star,
its slow expansion leads to a long phase of nuclear-timescale mass transfer (a few times $10^5$ yr,  $\mdot \sim 10^{-6}-10^{-5} \msunyr$). 
The slow mass transfer dominates the rest of the evolution: 
it persists at least until the end of core-He burning, at which point we terminate our models (the remaining lifetime 
is relatively short, $< 0.1$ Myr). Occasionally the mass exchange may be interrupted by temporary detachments, in particular 
around the age of $7.6$ ($7.7$) Myr of the LMC (SMC) models. These are
caused by (typically subtle) contractions of the donor star in response to changes of the hydrogen abundance in the moving location 
of the H-burning shell. \footnote{A similar transition in the H shell was previously found to trigger a blue-loop evolution in $15 \msun$ stellar models 
by \citet{Langer1985}.}
Oscillations in the mass transfer rate during the nuclear-timescale phase are associated with changes in the 
helium abundance of the outer donor layers as it is being stripped deeper into layers that used to be convective 
during MS (see Sec.\ref{sec:why_partstrip}).
The few low-Z models with the lowest envelope masses $M_{\rm env;left} \lesssim 2.5 \msun$ left after the initial rapid mass transfer, 
($q \lesssim 0.4$ at LMC and $q = 0.25$ at SMC composition) show a somewhat different type of evolution. 
These binaries remain detached for most of their He-burning lifetime, with possibly only a relatively brief episode
of nuclear-timescale mass transfer close to the end of this phase.

Notably, in none of the LMC or SMC models in Fig.~\ref{fig.diffQ} is the envelope fully lost in winds or mass transfer (by the end of core-He burning), which is partly due to weaker winds at low Z.
In fact, at no point does the primary become a hot ($\logteff \gtrsim 4.6$) stripped helium star, in sharp 
contrast to the Solar metallicity case. See Fig.~\ref{fig:square_env_remain} for the amount of envelope left at the end of core-He burning across all our models. 

\begin{figure}
\includegraphics[width=\columnwidth]{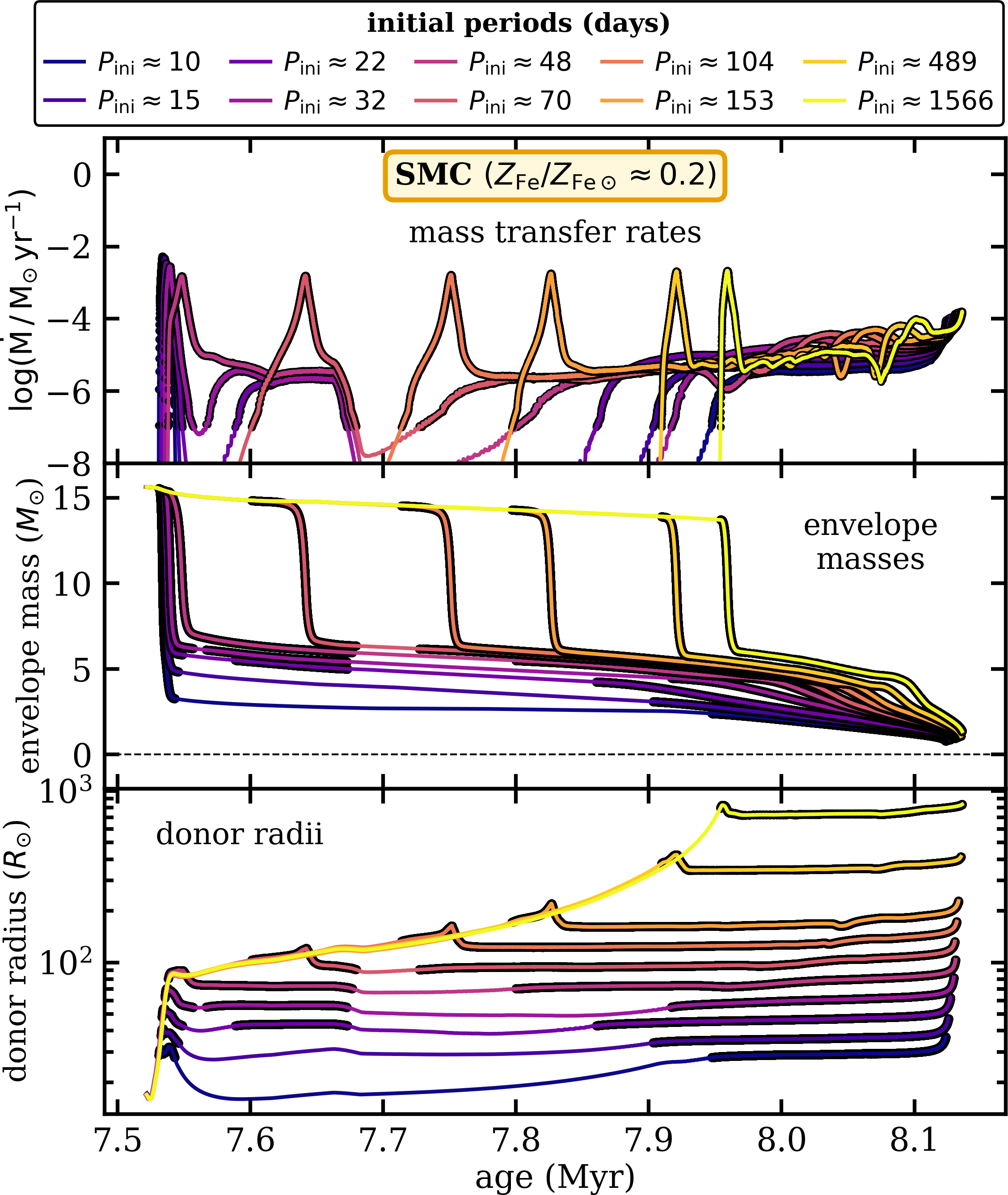}
\caption{Mass transfer evolution in binaries with a $25 \msun$ primary (initial mass),  SMC composition, and various initial 
orbital periods $P_{\rm ini}$. The shorter the orbital period, the lower the envelope mass $M_{\rm env;left}$ 
that remains after the initial phase of rapid (thermal) mass transfer.}
\label{fig:period_impact_SMC}

\end{figure}
Besides the mass ratio, another factor that affects the degree of envelope stripping at low Z
is the initial orbital period $P_{\rm ini}$. In Fig.~\ref{fig:period_impact_SMC}, we show this for the case of 
models with a $M_1 = 25 \msun$ primary, the mass ratio $q = 0.6$, and the SMC metallicity (see Fig.~\ref{fig:period_impact_LMC} for the LMC case).
As the top and the middle panel illustrate, the shorter the orbital period, the lower the mass of the remaining envelope after the initial rapid mass transfer phase ($M_{\rm env;left}$).
We notice that above a certain period (or equivalently mass ratio), the donors tend to slow down the mass transfer rate 
at similar $M_{\rm env;left}$ values and then continue to transfer 
mass on the nuclear timescale. In the case shown in Fig.~\ref{fig:period_impact_SMC}, this happens for $P_{\rm ini} \gtrsim 30$ days and leads to $M_{\rm env;left} \approx 6-7 \msun$. Notably, the donors in those models are already near the end of the HG expansion or more evolved at the onset of RLOF (see the botom panel).

As previously mentioned and illustrated in the donor radius panel of Fig.~\ref{fig:period_impact_SMC}, 
temporary detachments happening at $\sim 7.7$ Myr are caused by relatively modest contractions 
of donors during the core-He burning phase (by $10-20\%$ in this particular case).  

We discuss the above findings in view of the donor envelope structure in Sec.~\ref{sec:why_partstrip}.
We note that although not explored in the current study, the degree of envelope stripping is expected to also depend 
on the assumptions about the accretion efficiency and angular momentum of the non-accreted 
matter, i.e. all the factors that influence the evolution of the Roche lobe size of the donor star.


The reason why the mass transfer evolution is so different between the Solar and low-Z cases in Fig.~\ref{fig.diffQ} 
is related to radial expansion of mass-losing stars and the 
size that a partially-stripped donor of a given metallicity would need to have in order to regain thermal equilibrium. 
We discuss this is detail in Sec.~\ref{sec:why_partstrip} and Fig.~\ref{fig:MT_transition}.
In summary, we find that a partially-stripped donor of Solar $Z$ would need to expand to a size of a red supergiant to regain equilibrium, 
$\sim 1800\rsun$, which is much larger than a size of its Roche lobe. As a result, the donor keeps rapidly expanding and the thermal-timescale mass transfer 
continues until it strips nearly entire envelope, at which point the donor finally contracts to become a hot stripped star. 
In contrast, the donors of LMC and SMC compositions ($M_1 = 25 \msun$) that are partially-stripped in the initial phase of rapid mass transfer
can be in thermal equilibrium at a much smaller size $\sim 50\rsun$, 
which is similar to that of their Roche lobes. As a result, they can regain equilibrium and slow down the mass transfer interaction
or even detach.
In a similar way, single massive low-Z stars may remain relatively compact before the onset 
of the core-He burning, avoiding a rapid HG expansion until the red giant branch (in particular the $25 \msun$ models, 
see the radius-age diagram on the right-hand side of Fig.~\ref{fig.diffQ}).
In fact, as we will show in Sec.~\ref{sec:nuc_param_space}, we find nuclear-timescale mass transfer evolution 
in models with donor masses roughly from the mass range in which the halted HG expansion happens.

\begin{figure}
\includegraphics[width=\columnwidth]{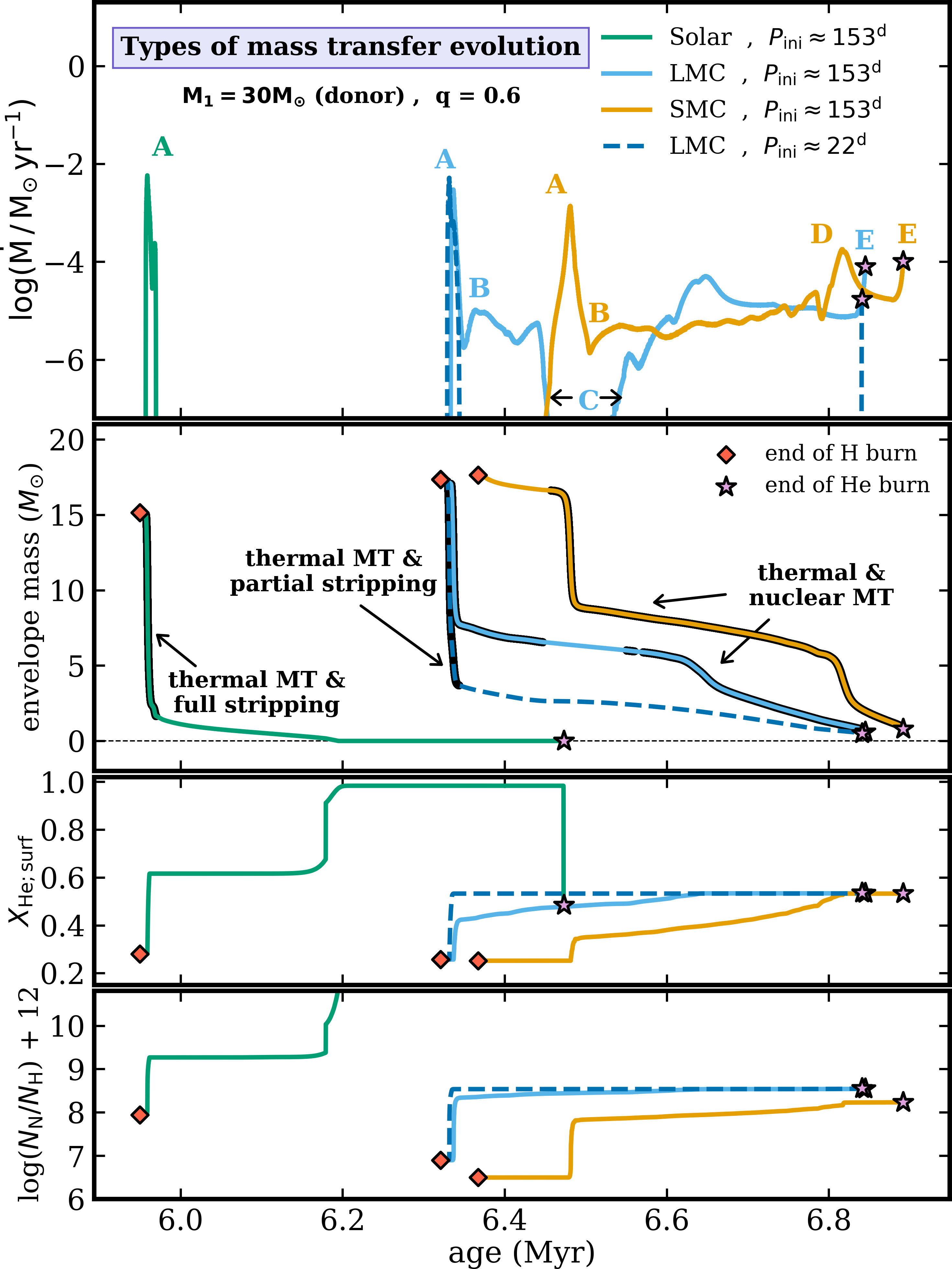}
\caption{Three different types of mass-transfer evolution found across our binary models, illustrated 
with cases with $M_1~=~30\msun$ primaries and different compositions (metallicities).  
Going from top to bottom, subsequent panels show: mass-transfer rate, envelope mass, surface He abundance, and surface nitrogen enrichment.
Bold parts of the curves in the second panel indicate phases of RLOF. 
Letters mark distinctive events that may occur during the evolution as following. A: a phase of rapid thermal-timescale mass transfer, B: a point when the donor regains thermal equilibrium and the mass transfer
transitions to the nuclear timescale, C: phases of detachment, D: temporary mass-transfer rate increase when
the donor becomes stripped down to layers that used to be in the intermediate convective zone,  
and E: mass transfer peak associated with the end of core-He burning and re-expansion of the donor.}
\label{fig:sns_30example}
\end{figure}

\begin{figure}
\includegraphics[width=\columnwidth]{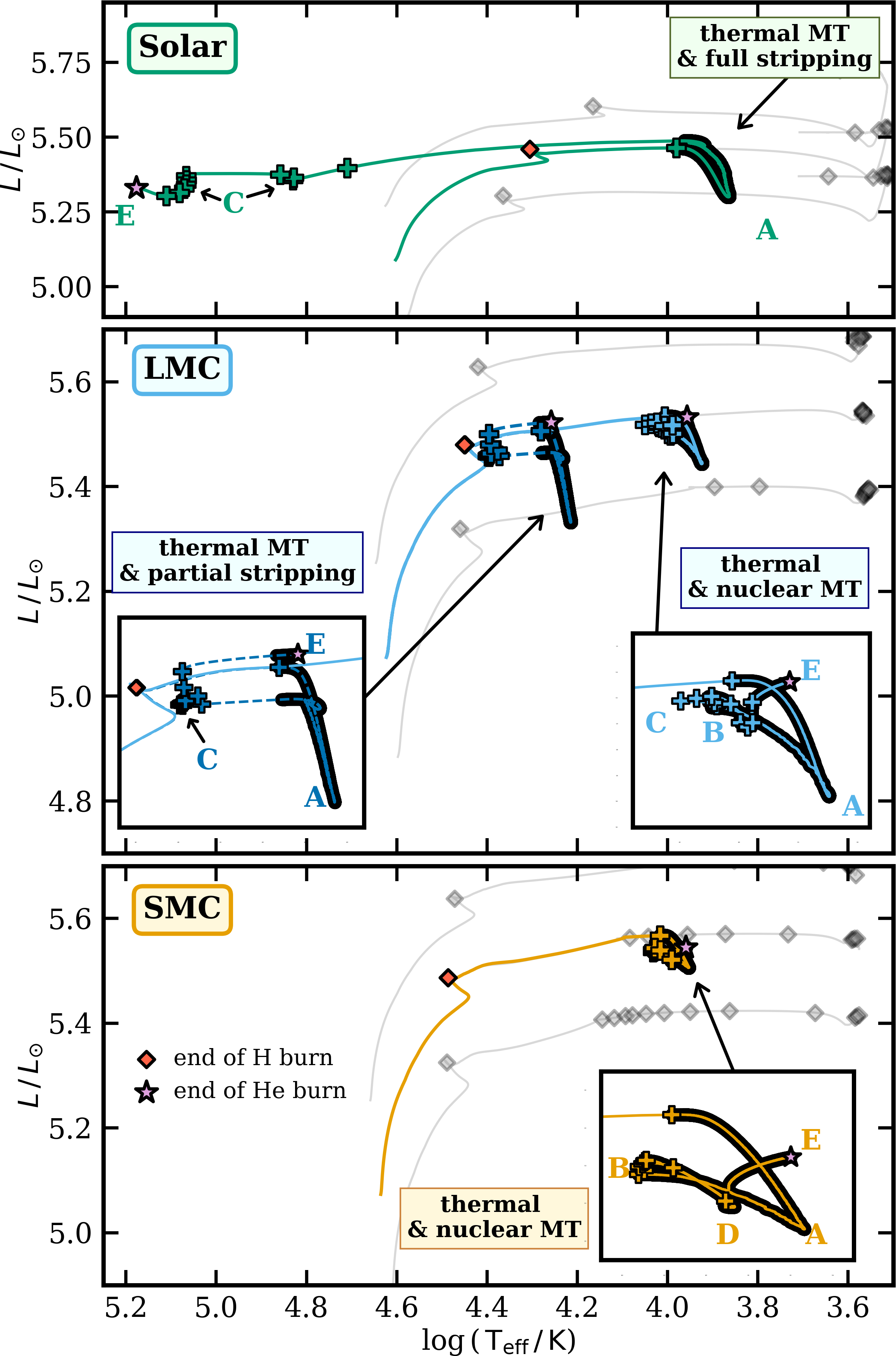}
\caption{Evolution of $30\msun$ primaries (donors) from models shown in Fig.~\ref{fig:sns_30example} in the HR diagram. 
Letters correspond to the same events in the mass transfer evolution as before (see caption of Fig.~\ref{fig:sns_30example}
or text). Diamond (star) symbols mark the end of core H (core He) burning. These models illustrate the three main types of stable post-MS mass transfer evolution: thermal-timescale 
mass transfer producing hot (fully) stripped stars (top panel), thermal-timescale mass transfer producing much cooler, partially 
stripped stars (middle panel, case on the left-hand side), as well as binaries that evolve through phases 
of both thermal- and nuclear-timescale mass transfer (the remaining two models). 
}
\label{fig:sns_30exampleHRD}
\end{figure}


\subsection{Three types of post-MS mass-transfer evolution}

Throughout our binary grid, we find three qualitatively different types of mass-transfer evolution, which we illustrate in Fig.~\ref{fig:sns_30example} (using the $M_1 = 30\msun$ case as an example).
In Fig.~\ref{fig:sns_30exampleHRD}, we plot the corresponding donor tracks in the HR diagram, 
with the phase of RLOF shown in bold and 
crosses indicating where the donor spends most of its core-He burning lifetime (spaced every 50,000 yr). 

The Solar metallicity donor transfers nearly its entire envelope ($M_{\rm env;left} \approx 1.7 \msun$) 
during a phase of thermal-timescale mass transfer 
(labeled 'A'). After detachment it rapidly contracts and moves leftwards in the HR diagram to spend most of 
its He-burning lifetime ($\sim 85\% \Delta t_{\rm He;burn}$) as a hot UV-bright stripped
star 
('C' in Fig.~\ref{fig:sns_30exampleHRD}) before central He exhaustion ('E' in Fig.~\ref{fig:sns_30exampleHRD}). For most of its lifetime, the stripped donor has an effective temperature of initially $\logteff \approx 4.85$ and then subsequently $\logteff \approx 5.1$. The transition to the higher $T_{\rm eff}$ takes place at the age of $\sim 6.2$ Myr. At that point the stellar wind removes the remaining part of the envelope with $X_{\rm He;surf} \approx 0.6$ and reveals deeper layers with a higher He abundance and a much steeper He/H gradient, leading to a further contraction (see Fig.~\ref{fig:abundance_profiles} for a few examples of He abundance profiles in stripped stars).

\begin{figure*}
\includegraphics[width=\textwidth]{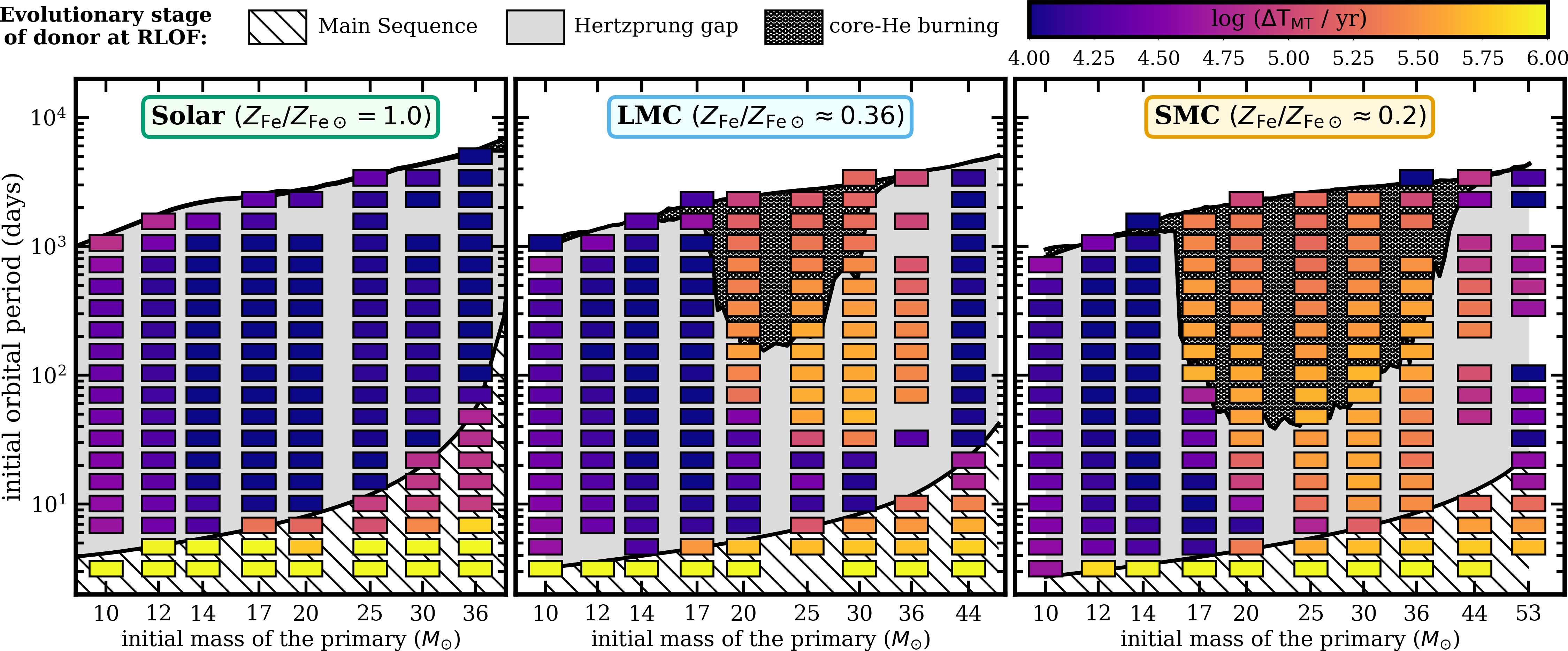}
\caption{Integrated duration of mass transfer through RLOF  ($\rm \Delta T_{\rm MT}$) in all the binary models computed
across different primary masses $M_1$, orbital periods $P_{\rm ini}$, and metallicities. Each binary evolution model is represented by a rectangle colored according to 
its corresponding $\rm log(\Delta T_{\rm MT} / yr)$ value. Missing rectangles are non-converged models. The binary models are mapped onto the parameter space of varying $M_1$ and $P_{\rm ini}$, 
with different background areas indicating the evolutionary state of the donor star at the moment of the initial RLOF (similar to Fig.~\ref{fig.bin_param_ranges}):
the MS donors (dashed), rapidly expanding HG donors (light grey), and core-He burning donors (densely dotted). 
Rectangles colored in various shades of yellow and orange correspond to models that evolve through a phase of nuclear-timescale 
mass transfer ($\rm log(\Delta T_{\rm MT} / yr) \gtrsim 5$)}
\label{fig:square_MTdur}
\end{figure*}

The LMC donor in a binary with $P_{\rm ini} \approx 22$ days (dashed line) also experiences 
only a brief phase of thermal-timescale mass exchange ('A'). In contrast to the Solar case, 
it retains a bigger envelope ($M_{\rm env;left} \approx 3.7 \msun$) that is never fully lost in winds. 
The donor burns He as a partially stripped star in a detached binary ('C'). 
It is much cooler and bigger than a stripped star: $\logteff \approx 4.4$ and $R \approx 30 \rsun$.
In fact, its locus in the HR diagram is similar to that of a single $30 \msun$ star at the end stages of MS or 
early stages of post-MS evolution. The post-interaction nature of a partially-stripped donor may be 
revealed through measurements of increased helium and nitrogen abundances (see the bottom two panels of Fig.~\ref{fig:sns_30example}) as well as its low spectroscopic mass,
see Sec.~\ref{sec:res_partstrip} for a further discussion. 
At the very end of core-He burning, the donor expands again leading to another phase of mass transfer 
(the so-called case BB RLOF, 'E'). The outcome of this phase and the final evolutionary stages before the core-collapse of the donor 
are outside the scope of this work. 

The evolution of the LMC model with $P_{\rm ini} \approx 153$ days as well as the SMC model is dominated 
by long phases of nuclear-timescale mass transfer. After the initial rapid mass-transfer (labeled 'A'), 
these donors regain thermal-equilibrium and the mass-transfer rate decreases ('B').
During the subsequent phase of slow and long-lived mass transfer, their position in the HR diagram hardly changes at all.
Note that the temporary detachment in the LMC model ('C') is associated with only a slight contraction 
of the donor star and therefore no significant displacement in the HR diagram. We find that binaries evolving 
through a nuclear-timescale mass transfer often experience a secondary ('delayed') peak of mass transfer rate 
which we label with 'D' in Fig.~\ref{fig:sns_30example} and Fig.~\ref{fig:sns_30exampleHRD}. 
This peak in $\dot{M}$ is associated with the donor being stripped down to a helium abundance plateau left by 
layers that used to be fully convective  
as part of the so-called intermediate convective zone. We discuss this in more detail in Sec.~\ref{sec:why_partstrip}.
Once again, at the final stages of He-burning the donor stars expand leading to a phase of thermal-timescale 
mass transfer at the point of central-He exhaustion ('E'). 

In the following sections we discuss the parameter space for the different types of mass transfer evolution illustrated above. 

\subsection{Parameter space for nuclear-timescale mass transfer}
\label{sec:nuc_param_space}

\begin{figure*}
\centering
\includegraphics[width=0.95\textwidth]{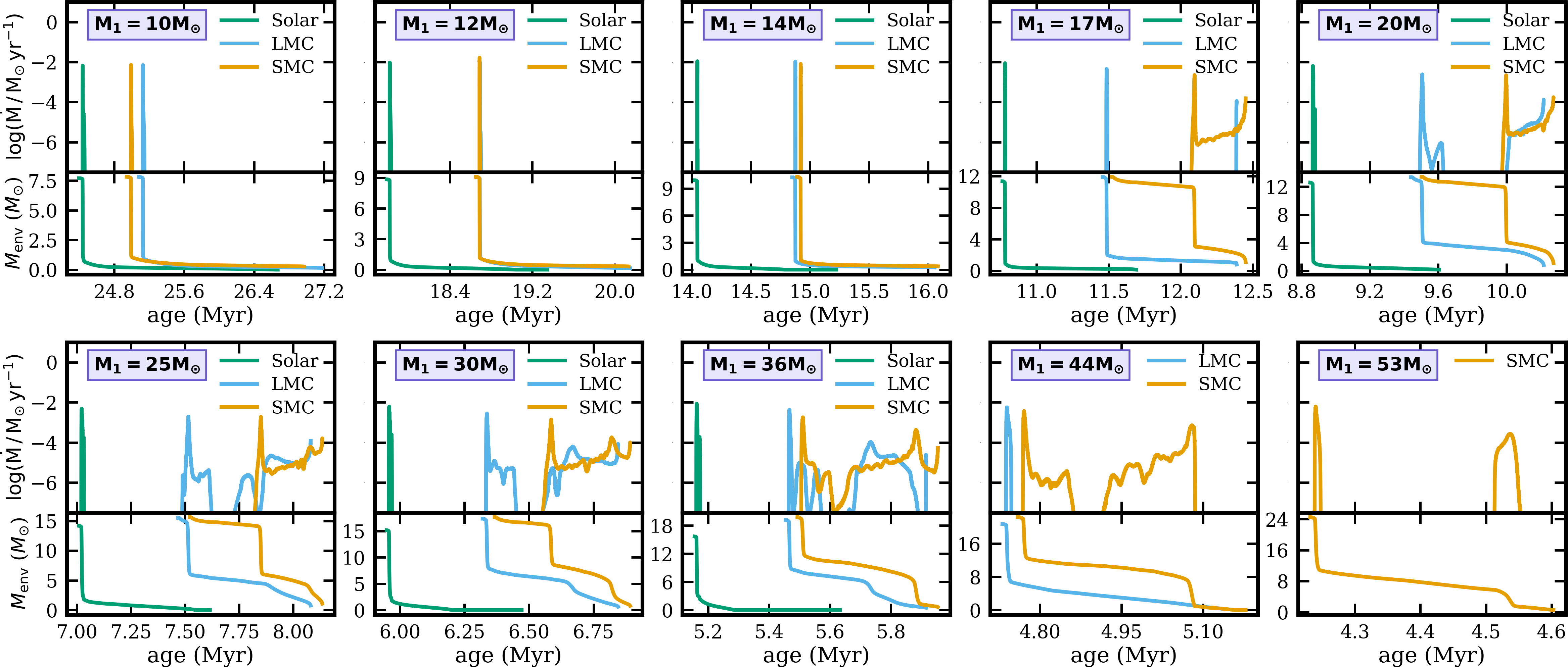}
\caption{Mass-transfer rate and the envelope mass as a function of time in binary models evolving through 
a case B mass-transfer. Different panels correspond to different initial masses of the donor ($M_1$). Colors 
indicate different metallicities. 
All the models were computed with initial orbital period $P\approx 225$ days and initial mass ratio $q = 0.6$. 
At Solar metallicity, the case B mass transfer is always a short-lived phase ($<10^4$ yr) that strips the (nearly) entire 
hydrogen envelope leaving behind a stripped helium core. At subsolar metallicities of the LMC and SMC, above a certain donor mass
a significant part of the envelope may be stripped on a much longer nuclear timescale of core He burning. 
Such donors remain only partially stripped for the duration of core-He burning. }
\label{fig:diffM1}
\end{figure*}

\begin{figure*}
\centering
\includegraphics[width=0.95\textwidth]{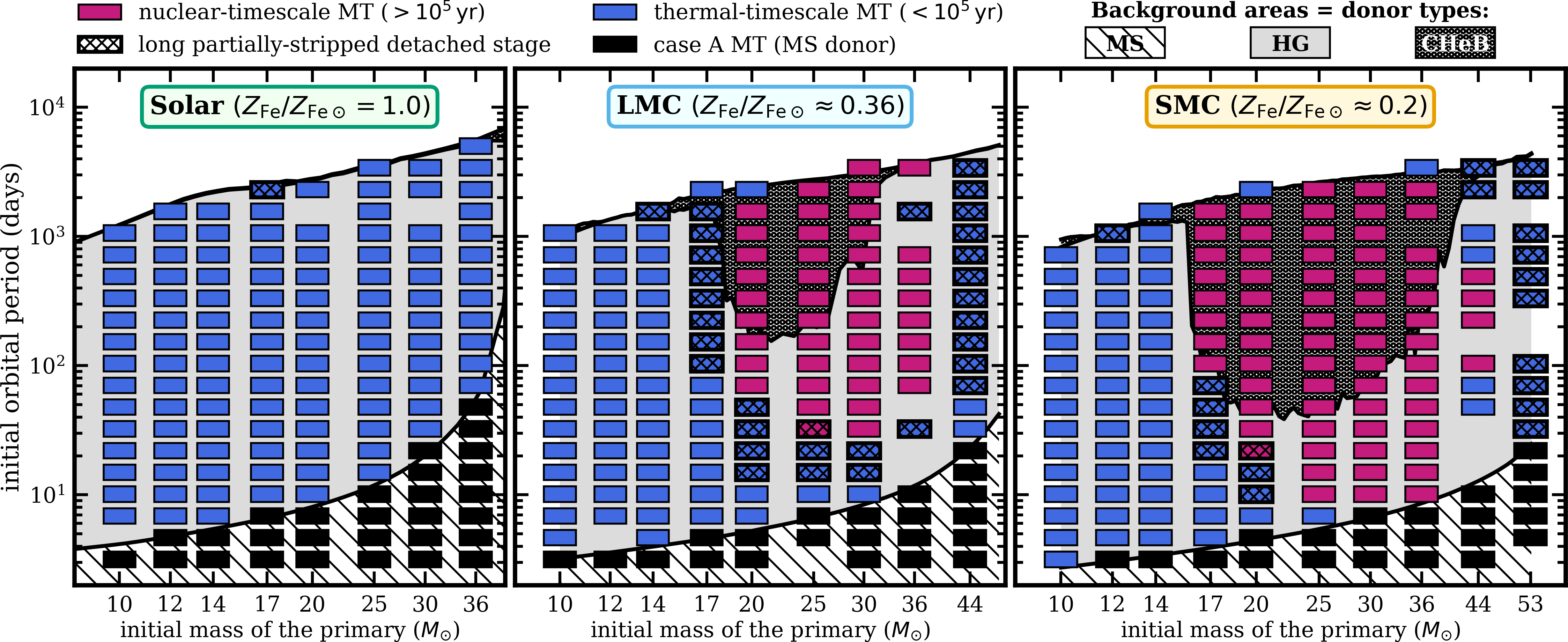}
\caption{Major types of mass transfer evolution found across the grid of binary models. Similarly to Fig.~\ref{fig:square_MTdur}, 
each binary model is represented by a rectangle and the background areas indicate the evolutionary state of the donor at the moment of the initial 
RLOF (Main Sequence, Hertzprung gap, or core-He burning). Missing rectangles are non-converged models. Black rectangles correspond 
to models in which the mass transfer takes place already during the MS. Among binaries with a post-MS interaction, we differentiate between 
those that evolve through phase(s) of nuclear-timescale mass transfer ($\rm \Delta T_{\rm MT} > 10^5$ yr, in magenta) and those 
in which the interaction happens on a short thermal timescale ($\rm \Delta T_{\rm MT} < 10^5$ yr, typically  $< 10^4$ yr, in blue). 
In addition, in hatch are models in which the post-interaction donor phase with $T_{\rm eff} < T_{\rm eff;ZAMS}$ is long 
($> 0.75$ the core-He burning lifetime), i.e. models leading to partially-stripped rather 
than hot stripped stars.}
\label{fig:tricolor}
\end{figure*}

In the previous section we demonstrated that a post-MS mass transfer 
in massive low-Z binaries can be long and slow, proceeding on a nuclear-timescale and leading to only partial 
envelope stripping. In this section we analyze the parameter space for this type of evolution. 
Fig.~\ref{fig:square_MTdur} includes results from our entire grid of binary models (initial mass ratio $q = 0.6$). 
Each model is represented by a rectangle that is colored according to the integrated duration of RLOF 
mass transfer in that model, on a logarithmic scale from $10^4$ to $10^6$ yr. 
Non-interacting models (i.e. the widest binaries) are not plotted. Empty spaces without a rectangle indicate non-converged models (due to numerical issues).
In addition, similarly to Fig.~\ref{fig.bin_param_ranges}, in the background of Fig.~\ref{fig:square_MTdur} we indicate  
the parameter ranges for RLOF corresponding to different evolutionary stages of the donor star: 
MS donors (dashed), rapidly-expanding HG donors (light grey), and core-He burning donors (densely dotted). 
Note that each mass includes at least one case-A mass transfer model, i.e. with a donor star that is still on the MS. 
In what follows we focus only on wider binaries: those that interact during the post-MS evolution of the donor.

Rectangles colored in various shades of orange in Fig.~\ref{fig:square_MTdur} correspond to models that evolve through a phase of nuclear-timescale 
mass transfer (with duration $\rm log(\Delta T_{\rm MT} / yr) > 5$). 
In
a similar Fig.~\ref{fig:tricolor}, which maps the major types of mass 
transfer evolution found across the grid of models, we clearly 
distinguish the nuclear-timescale cases in magenta.
We find that the parameter space for this type of evolution 
is related 
to the parameter space for RLOF from core-He burning donors, but not identical. This is the mass range in which primaries experience a low degree 
of rapid post-MS expansion and burn He as blue or yellow supergiants. 
Intuitively, it makes sense that those are the donors that can regain thermal equilibrium while only partially 
stripped and transfer mass on the nuclear timescale of core-He burning.
Somewhat surprisingly, we find slow mass transfer and partial-envelope stripping also in various models in which the donor at RLOF is still a rapidly-expanding HG star. For example: models with $P_{\rm ini} \approx 10-30$ days and $M_1 = 20-36 \msun$ in the SMC grid and models with $M_1 = 36 \msun$ and $M_1 = 44 \msun$ donors in the LMC and SMC cases, respectively.
Notably, in all these cases the detachment or nuclear-timescale mass transfer occurs only after the core-He burning phase is reached. 
Fig.~\ref{fig:tricolor} illustrates that the parameter space for nuclear-timescale mass transfer (and partial-envelope stripping in general) is significantly larger than simply the parameter space for mass transfer initiated at the core-He burning stage.

\begin{figure*}[h]
\includegraphics[width=\textwidth]{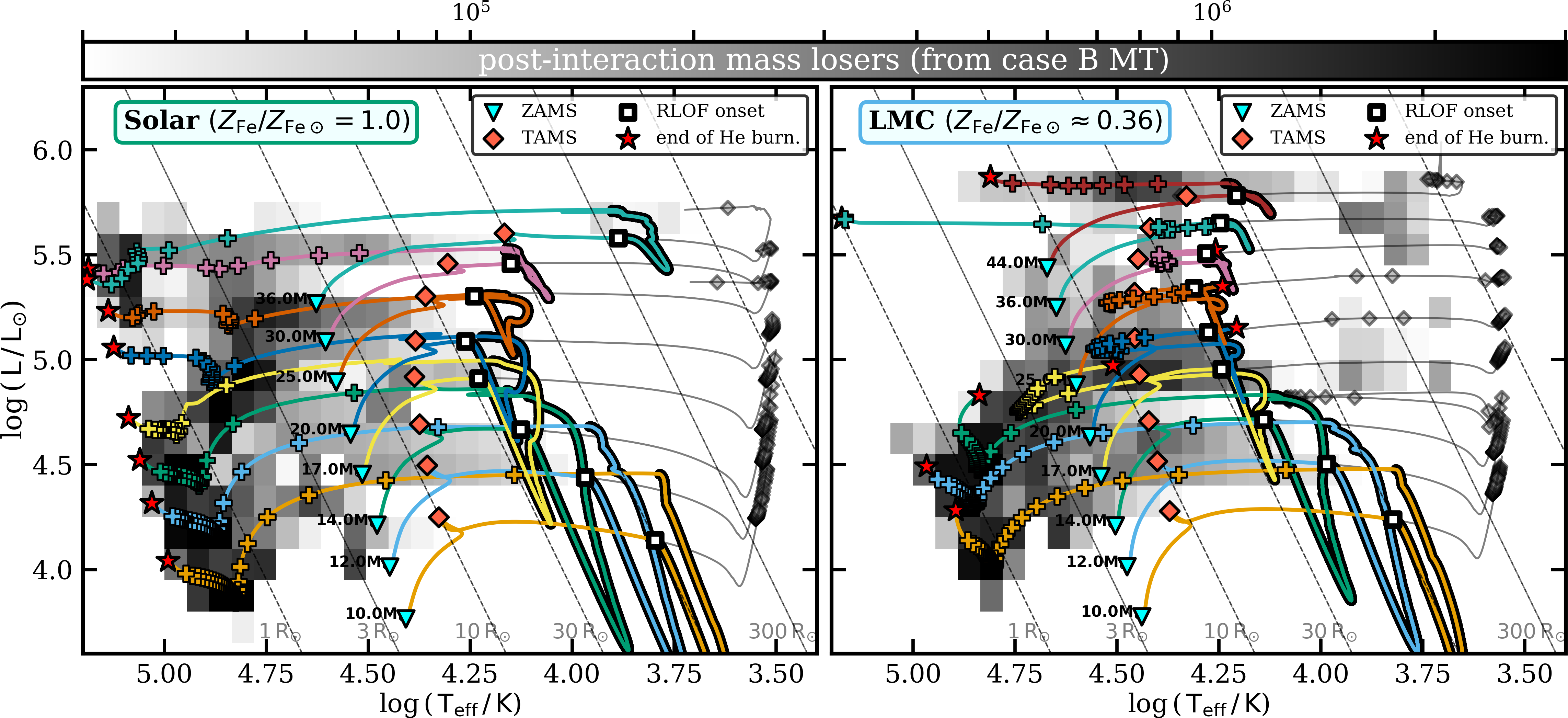}
\caption{Distribution of post-interaction donors in detached binaries with Roche-lobe filling factors $< 0.6$ (shown in greyscale)
from all the grid models interacting after the end of MS, i.e. through case B mass transfer, plotted 
for the Solar metallicity (left) and the LMC composition (right). Overplotted are several donor tracks from selected binary models (colored lines) as well as single stellar tracks for comparison (grey lines).
The mass transfer phase is shown in bold. Crosses are spaced by 50,000 yr starting from the onset of RLOF. 
Note that models were not weighted by the initial-mass function or the initial period distribution 
to construct the distribution. In all the Solar metallicity models, thermal-timescale mass transfer 
leads to the formation of hot stripped stars. In the LMC models, primaries above the initial mass of $M_1 \geq 17\msun$ 
become only partially stripped and as post-interaction stars can largely overlap with single-stellar tracks. 
}
\label{fig:HRD_PShist2d}
\end{figure*}

The effect of the primary mass and metallicity on the mass transfer duration is also evident in Fig.~\ref{fig:diffM1}, 
in which we take a slice through the grid of binary models selecting those with initial periods of $P_{\rm ini} \approx 225$ days 
and varying the initial primary mass $M_1$. Above a certain primary mass, 
the low-Z donors retain a substantially larger fraction of their envelopes 
from the initial rapid interaction and experience episodes of slow mass transfer afterwards. 
Notably, LMC systems with $M_1 = 17\msun$ and $44 \msun$ as well as the SMC model with $M_1 = 53 \msun$ 
evolve through mostly rapid (thermal) mass transfer and yet they retain a higher envelope fraction 
compared to Solar metallicity models. Those model lead to the formation of long-lived partially stripped 
stars in detached binaries, similarly to the $M_1 = 30 \msun$ LMC model with $P_{\rm ini} \approx 22$ days
shown earlier in detail in Fig.~\ref{fig:sns_30example} and Fig.~\ref{fig:sns_30exampleHRD}.

\subsection{Formation and properties of partially-stripped stars in detached binaries}

\label{sec:res_partstrip}

In this section we take a closer look at models that lead to formation 
of long-lived partially-stripped stars in detached binaries 
(e.g. Fig.~\ref{fig:sns_30example} and Fig.~\ref{fig:sns_30exampleHRD}). 
We define those as binaries in which a donor star spends at least 
$75\%$ of its core-He burning lifetime as a post-interaction star that stays on the cooler side of the ZAMS temperature in the HR diagram. Such donors are partially-stripped 
of their envelopes and overlap with single stellar tracks, as opposed to fully stripped stars that become hot helium 
stars (with $T_{\rm eff} \gg T_{\rm ZAMS}$), see Fig.~\ref{fig:sns_30exampleHRD}. 

In Fig.~\ref{fig:tricolor}, with hatches we mark which models in our grid lead to partially-stripped stars.
It should be noted that various non-hatched magenta rectangles (representing models that evolve through nuclear-timescale mass transfer)
can also produce partially-stripped donors but only for relatively short detachment phases (that 
do not make our $0.75 \Delta T_{\rm He;burn}$ cut). For more details see Fig.~\ref{fig:cooldetach}
in the appendix. Importantly, in those cases the donor star 
never shrinks significantly with respect to its Roche lobe (usually $R_{\rm don} / R_{\rm RL} > 0.6$).
\footnote{The detachment phases in nuclear-timescale mass transfer models are associated with changes in the H abundance 
in the moving location of the H-burning shell. Given the uncertainties in the detailed features of 
the H abundance profile of the bottom envelope layers, these detachments are not a robust 
prediction of evolutionary models.}

Fig.~\ref{fig:tricolor} makes it clear that in the $M_1-P_{\rm ini}$ plane
models leading to partially- stripped stars are an intermediate stage between 
rapid mass transfer models and models experiencing nuclear-timescale 
mass transfer. 
Similarly to nuclear-timescale interaction models, we find that 
partial-envelope stripping is roughly associated with the primary mass range in which the HG expansion is halted 
at the blue/yellow supergiant stage. 
Additionally, for a given primary mass among the LMC and SMC models with $M_1 \geq 17 \msun$,
the type of mass transfer evolution depends on the orbital period, where 
long detachments with partially-stripped stars usually take place in binaries with relatively short
orbital periods. This is because the shorter the orbital period, the more of the envelope 
is transferred in the initial mass transfer phase (which favors long detachments), see also Fig.~\ref{fig:period_impact_SMC}. 
A similar effect would be observed in models with a more extreme mass ratio or a higher specific angular momentum of
the non-accreted matter. We also find relatively fewer
binaries with long detachment phases among the SMC models compared to the LMC grid.

In the following, we discuss the basic properties of partially-stripped stars. 
Fig.~\ref{fig:HRD_PShist2d} compares the distribution of donors stripped in the LMC metallicity models 
with their Solar metallicity counterparts in the HR diagram. In each panel, the coloured lines show donor tracks from a selection 
of binary models with different primary masses in which the mass transfer was initiated soon after the end of MS. 
Crosses indicate where a donor spends most of its core-He burning lifetime since the onset of RLOF (spaced by 50,000 yr). 
Grey lines show single-stellar models of the same masses. In addition, in greyscale we plot the distribution 
of post-interaction donors in a detached state with a Roche-lobe filling factor $R/R_{\rm RL} < 0.6$
from all our binary models. The 0.6 cut makes sure that we do not include most of binaries that evolve primarily through phases 
of nuclear-timescale mass transfer (and temporary detachments with $R/R_{\rm RL} > 0.6$) 
with blue or yellow supergiant donors at $\logteff < 4.0$. We discuss those in Sec.~\ref{sec:disc_CHEB_HRD}.
We caution that the binary models used to construct 
the greyscale distribution were not weighted by either the initial mass function or the initial period distribution
(for a weighted HRD distribution we refer to Fig.~\ref{fig:disc_HRD_hist2d}).

\begin{figure}
\includegraphics[width=\columnwidth]{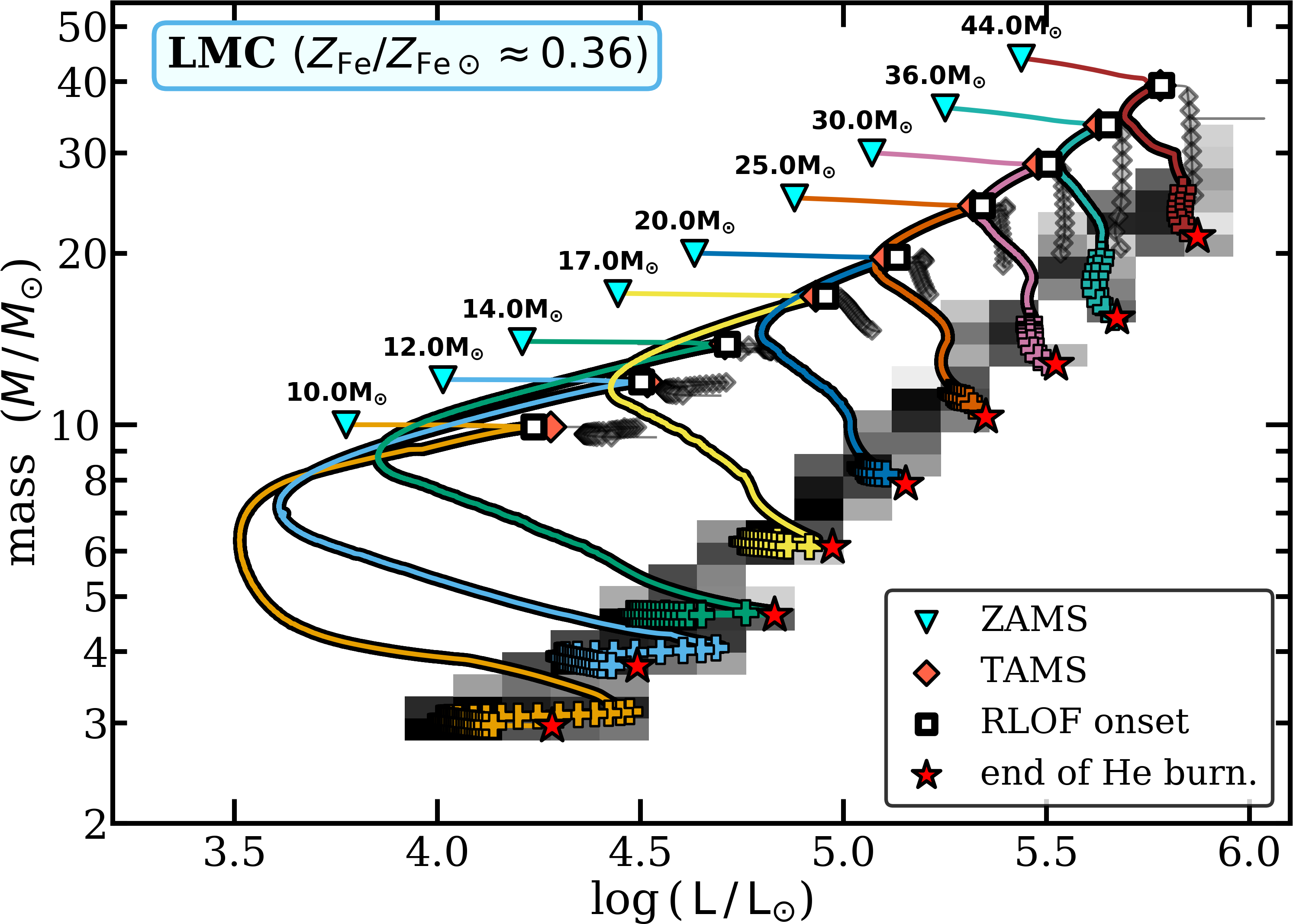}
\caption{Luminosity-mass relation of post-interaction donor stars in detached binaries 
with $R/R_{\rm RL} < 0.6$ at the LMC metallicity. The colors correspond to the few selected donor tracks 
from binary models shown in the HR diagram in Fig.~\ref{fig:HRD_PShist2d}, with crosses (spaced by 50,000 yr) marking 
the position occupied for most of the evolution since the onset of RLOF. Grey lines show the  evolution of single 
stars with the corresponding masses. Similarly to Fig.~\ref{fig:HRD_PShist2d}, in greyscale we plot the distribution 
of post-interaction stars in detached systems from all the grid models (except those interacting on MS). The figure demonstrates 
that mass transfer can produce partially-stripped stars that may appear significantly undermassive for their apparent evolutionary 
mass (as one would deduce from single tracks). Note, however, that models with $M_1 \leq 14 \msun$ produce post-interaction 
stars that are hot stripped stars ($T_{\rm eff} > T_{\rm eff;ZAMS}$).}
\label{fig:PS_masses_LMC}
\end{figure}

At Solar metallicity (left panel in Fig.~\ref{fig:HRD_PShist2d}), all the binary models evolve through 
thermal-timescale mass transfer which rapidly strips nearly the entire envelope of the donor. As a result, 
most of the post-interaction lifetime of donors is spend in the hot region of stripped helium stars, to the left 
of the $T_{\rm eff}$ range of ZAMS. 
It is noteworthy that hot stripped stars originating from $\geq 17\msun$ donors (luminosities $\logL \gtrsim 4.9$) tend to be somewhat cooler than those originating from lower mass stars. This is because, in general, more massive donors tend to retain a more massive envelope after mass transfer. As a result, above a certain donor mass ($\sim 17-20\msun$ in our grid), stripped stars still retain some of the envelope layers characterized by a relatively low He abundance ($Y \sim 0.6$) and a nearly flat abundance profile, see Fig.~\ref{fig:abundance_profiles}. Such remnant envelopes lead to stripped stars with cooler effective temperatures compared to those stripped more deeply into layers with a steep He/H abundance gradient \citep[see also][]{Schootemeijer2018}. This effect is present until luminosity $\logL \approx 5.3$, above which stellar winds are strong enough to quickly strip these remaining envelope layers and increase the effective temperature to $\logteff > 5.0$.

In the LMC case, models with primaries with $M_1 \geq 17 \msun$ and a post-MS mass-transfer interaction no longer produce 
hot stripped stars. Instead, they lead to partially-stripped post-interaction donors that populate the HR diagram 
region with $T_{\rm eff} < T_{\rm eff;ZAMS}$ in which they occupy the same region as pre-interaction stars or single stars.
Interestingly, we not only find post-interaction core-He burning stars in the region of MS (where, for most realistic star formation histories,
they would be vastly outnumbered by core-H burning stars) but also in the $T_{\rm eff}$ range of the early post-MS stage ($\logteff$ 
between $\sim 4.2$ and $4.4$), where single stellar tracks predict virtually not stars and a sharp gap in the HR diagram distribution, 
in contrast to observations (see Sec.~\ref{sec:disc_CHEB_HRD}).

\begin{figure}
\includegraphics[width=\columnwidth]{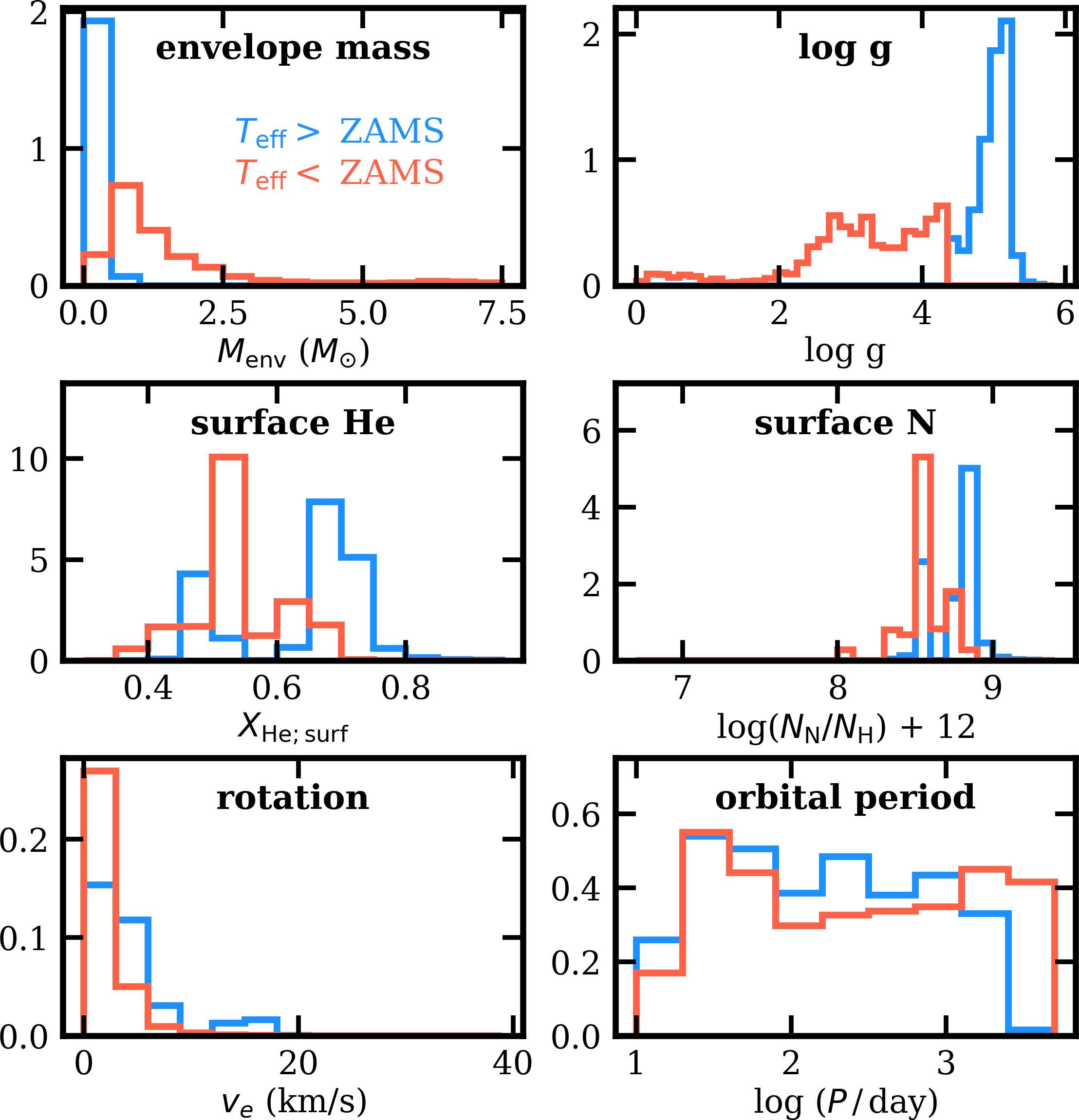}
\caption{Properties of hot stripped stars ($T_{\rm eff} > T_{\rm eff;ZAMS}$, in blue)
compared to those of partially-stripped ($T_{\rm eff} < T_{\rm eff;ZAMS}$, in red) 
in detached binaries with $R/R_{\rm RL} < 0.6$ (as in Fig.~\ref{fig:HRD_PShist2d}) based on LMC models 
with a post-MS mass transfer interaction (i.e. excluding case A models). Note that partially-stripped stars 
originate primarily from models with primary masses $M_1 > 17 \msun$.}
\label{fig:PS_histograms_LMC}
\end{figure}

Partially-stripped stars, while potentially mixed with single or pre-interaction stars, have
several distinctive characteristics. First, they are undermassive for 
they apparent luminosities compared to single stellar tracks. We illustrate this in Fig.~\ref{fig:PS_masses_LMC} where, 
similarly to Fig.~\ref{fig:HRD_PShist2d}, 
in greyscale we plot the mass-luminosity distribution of post-interaction stars in detached binaries from our entire grid of
LMC models (excluding case A systems). With colored lines, we overplot donor tracks from a few selected binary models (the same as in Fig.~\ref{fig:HRD_PShist2d}), 
and with grey lines we show single stellar tracks for comparison. 
The figure demonstrates 
that mass transfer can produce partially-stripped stars that may appear significantly undermassive for their apparent evolutionary 
mass (as one would deduce from their luminosity and based on single tracks). Wind-mass loss can reduce the mass of post-MS single stars in a similar way 
only for the most massive stars in our grid ($\gtrsim 44 \msun$).

In Fig.~\ref{fig:PS_histograms_LMC}, we illustrate several other basic properties of partially-stripped stars (in red) 
compared to hot stripped stars (in blue). In addition to their high luminosity-to-mass ratio, partially-stripped stars are also expected to be helium and 
nitrogen enriched on the surface, as prior mass loss has revealed deep envelope layers which have been 
mixed with products of CNO burning. At the same time, they are not expected to be fast rotators: they originate from wide 
binaries ($P_{\rm ini} > 10$ days) so that tidal synchronization 
leads to rotational velocities $\lesssim 10$ km/s.
As we find them all across the HR diagram, they have surface gravities in a wide range, from $\sim 2$ to $4.3$ (in log(g)), 
and can reside in systems with very different orbital periods (from tens to thousands of days).

In Sec.~\ref{sec:disc_CHEB_HRD} we discuss implications of our findings on the distribution 
and observables of core-He burning stars in the HR diagram.

\section{Why partially stripped? Radial response of stars to mass loss}

\label{sec:why_partstrip}

In the previous sections we found that in many low-Z massive binaries 
the post-MS mass transfer may include phases of long nuclear-timescale 
mass exchange ($\mdot \sim 10^{-5}\msunyr$, $\Delta T_{\rm MT} > 10^5 \, \rm yr$) 
and produce post-interaction core-He burning supergiants that are only partially stripped of their 
envelopes. This is in stark contrast to Solar metallicity models, 
all of which predict the thermal-timescale mass transfer 
to strip nearly the entire donor envelope ($\mdot \sim 10^{-3}\msunyr$, $\Delta T_{\rm MT} \lesssim 10^4 \, \rm yr$)
and produce hot stripped stars. 
In this section, we seek to understand the origin of these canonically different types of post-MS mass transfer evolution. 
To do so, we study the behavior of donor stars of different metallicities in response to mass loss. 

We define $R_{\rm th;eq}$ as the radius that a star needs to have in order to be in thermal equilibrium. In the case of singe-star evolution, 
most of the time $R_{\rm th;eq} = R$ where $R$ is the actual radius of a star.
The exception are short-lived phases such as the HG expansion or a phase of He-shell burning right after the central He exhaustion. 
In the case of binary evolution, the radius of a star is restricted by the size of its Roche lobe ($R \lesssim R_{\rm RL}$), 
which may prevent it from reaching thermal equilibrium. For instance, during thermal-timescale mass transfer the donor 
star is thermally unstable having radius $R_{\rm don} \approx R_{\rm RL} << R_{\rm th;eq}$. Its continuous fast expansion in a futile attempt
to equate $R_{\rm don} = R_{\rm th;eq}$ is what leads to high mass transfer rates.

\begin{figure}
\includegraphics[width=\columnwidth]{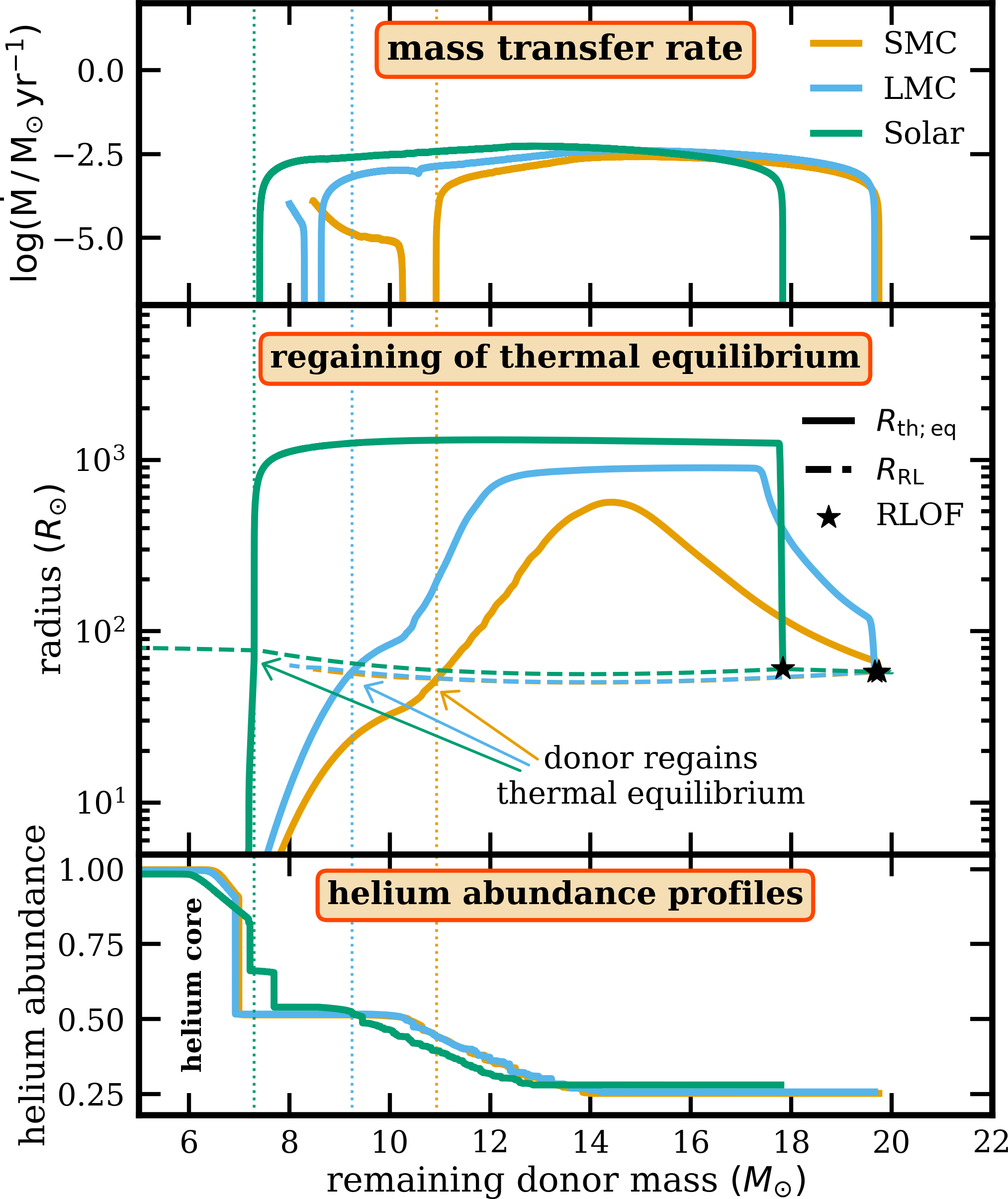}
\caption{Origin of partial envelope stripping of the LMC and SMC models 
explained based on the response of donor stars to mass loss, using the $20 \msun$ 
donor case as an example. 
Quantities are plotted as a function of the remaining donor mass $M_{\rm don}$, i.e. 
the time direction in the top two panels is to the left. Top panel: mass transfer rate 
from binary models ($P_{\rm ini} \approx 32$ days). Middle panel: 
equilibrium radius $R_{\rm th;eq}$ that the donor star would need to have in order to be in thermal
equilibrium, compared with its Roche lobe size $R_{\rm RL}$. For as long as $R_{\rm th;eq} > R_{\rm RL}$, 
the thermal-timescale mass transfer continues to strip the envelope of the donor. Vertical 
dotted lines mark the points when $R_{\rm th;eq} = R_{\rm RL}$, the donor can regain equilibrium, and the 
mass transfer may stop or slow down to the nuclear timescale. 
Bottom panel: internal He abundance profile of the donor at the onset of RLOF. 
}
\label{fig:MT_transition}
\end{figure}

Knowing $R_{\rm th;eq}$ 
of a donor star and how it changes over the course of mass transfer is the key to understand the transition 
from thermal-timescale mass transfer to detachment or to a nuclear-timescale mass exchange found in our SMC and LMC models. 
To obtain the evolution of $R_{\rm th;eq}$ as a function of the remaining donor mass $M_{\rm don}$, we proceed in
a similar way to \citet{Quast2019}. We begin by evolving a binary system until the onset of RLOF. Normally, from that point onward, 
all of our binary models would enter a phase of thermal-timescale mass transfer (at least initially). Here, however, we 
terminate the binary model, extract the donor star, and subject it to the following mass-stripping experiment. 
We switch off stellar winds and instead apply a constant mass-loss rate $10^{-5} \msunyr$ which is low enough to guarantee that the star remains at thermal equilibrium.
At the same time, we switch off any composition changes due to nuclear burning.
This mimics the rapid evolution through a phase of thermal-timescale mass transfer when there is no time for any significant burning to occur. 
The radius of a star that is being stripped in this way is a good representation of the equilibrium radius $R_{\rm th;eq}$
that a donor star would need to have in order to regain stability and stop the thermal-timescale mass transfer. 

In Fig.~\ref{fig:MT_transition}, we plot the results for the case of $20 \msun$ donors in binaries with an initial 
period of $P_{\rm ini} \approx 32$ days and three different metallicities. 
By the onset of RLOF ($R_{\rm RLOF} \approx 60 \rsun$), the SMC donor has nearly regained equilibrium from the HG phase as a core-He burning star, the LMC donor 
would still continue the HG expansion up to $R \approx 100 \rsun$ and the Solar donor up to $R \approx 1000 \rsun$.
The panels are plotted as a function of the remaining 
donor mass, i.e. the time direction in the top and middle panel is to the left. 
The bottom panel shows the internal helium abundance ($X_{\rm He}$) profile 
of the donor at the onset of RLOF. The He profiles are characterized by a plateau with a constant He abundance 
$X_{\rm He} \approx 0.5$ left by an intermediate convective zone and a constant $X_{\rm He}$ 
gradient above the plateau left by a retreating convective core during the MS \citep[e.g.][]{Langer1985,Langer1991}.
The middle panel shows the evolution of $R_{\rm th;eq}$ obtained 
from the mass-stripping experiment described above as well as the size of the donor's Roche lobe $R_{\rm RL}$ 
in the corresponding binary model (initial mass ratio $q = 0.6$). 
For as long as $R_{\rm th;eq} > R_{\rm RL} \approx R_{\rm don}$, the donor star is out-of-equilibrium and 
the mass transfer proceeds on the thermal timescale. Vertical dotted lines mark the donor mass when $R_{\rm th;eq} = R_{\rm RL}$.
For comparison, in the top panel we plot the actual mass transfer rate evolution 
from the corresponding binary models. Note that beside the initial thermal phase with $\mdot \approx 10^{-2.5} \msunyr$, 
the SMC model also experiences a nuclear-timescale mass transfer with  $\mdot \approx 10^{-5} \msunyr$.

The middle panel of Fig.~\ref{fig:MT_transition} illustrates the key difference between the Solar-metallicity and the low-Z donors. 
All throughout the mass transfer, the Solar-metallicity donor would need to expand to $R_{\rm th;eq} \approx 1000 \rsun$ (the size of a red 
supergiant) to regain thermal equilibrium. This is much larger than the size of the Roche lobe, not only in the $P_{\rm ini} \approx 32$ days 
and $q = 0.6$ example shown in Fig.~\ref{fig:MT_transition} but in nearly all the possible orbital configurations in an interacting binary. 
Only when the donor is stripped of nearly its entire envelope ($M_{\rm don} \approx 7.5 \msun$), does the $R_{\rm th;eq}$ begin to rapidly decrease 
towards the typical size of a fully-stripped helium star (a few $\rsun$) and the binary detaches. The $R_{\rm th;eq}$ evolution of 
the Solar model in Fig.~\ref{fig:MT_transition} is very well representative for all the post-MS donor stars that become fully stripped 
during thermal-timescale mass transfer, i.e. the entire Solar-metallicity grid as well as the $10$-$14 \msun$ donors at SMC and LMC compositions. 

The SMC and LMC donors in Fig.~\ref{fig:MT_transition} behave in a distinctively different way. The initial increase of $R_{\rm th;eq}$
is slower, which leads to slightly smaller thermal $\dot{M}$ rates found in low-Z compared to Solar binary models. More importantly, 
$R_{\rm th;eq}$ begins to decrease much earlier, when the donors are still far from being fully stripped.
\footnote{In fact, the $R_{\rm th;eq}$ behavior of low-Z models in Fig.~\ref{fig:MT_transition} is somewhat similar to that of MS donors.}
This allows the low-Z donors to regain thermal equilibrium and detach when only partially stripped, 
as seen in the binary models in the top panel (and found all throughout the low-Z grids of binary models). 
In addition, the fact that $R_{\rm th;eq}$ decreases gradually over a wide range of remaining donor masses is what leads
to a large variety of envelope masses of partially stripped low-Z donors, depending on the orbital period and the mass ratio
(e.g. Fig.~\ref{fig.diffQ} and Fig.~\ref{fig:period_impact_SMC}).

One may notice that the moment when $R_{\rm th;eq} = R_{\rm RL}$ in the mass-stripping experiment is not always exactly aligned with the $\dot{M}$
drop in the binary model (top panel), e.g. the LMC model in Fig.~\ref{fig:MT_transition}. This is because as the donor star relaxes 
to regain equilibrium at $R = R_{\rm th;eq}$, it may still continue to expand relative to its Roche lobe, leading to 'over-stripping'.
We find that this is especially the case when the $R_{\rm th;eq} = R_{\rm RL}$ vertical line falls in
the region of the He abundance plateau ($M_{\rm don}$ range between $\sim 7$ and $10 \msun$ in the SMC and LMC models in Fig.~\ref{fig:MT_transition}).

\begin{figure}
\includegraphics[width=\columnwidth]{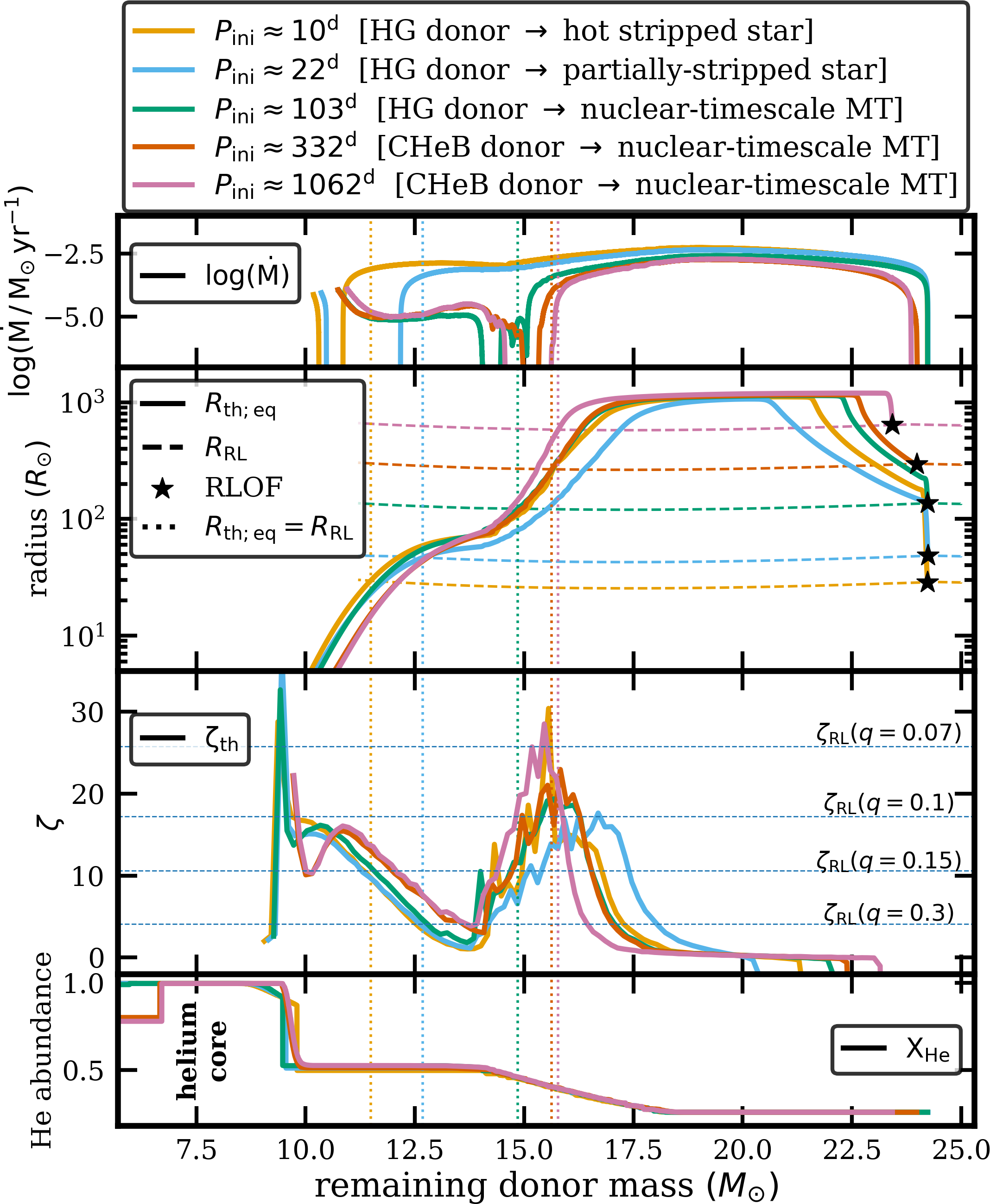}	
\caption{Similar to Fig.~\ref{fig:MT_transition}, but for the case of a $25 \msun$ donor at LMC composition 
and five different initial orbital periods. The legend on top details what was the evolutionary state 
of the donor at the onset of RLOF (HG phase or core-He burning) as well as the main outcome of the mass transfer 
interaction (hot stripped star, partially stripped star, or nuclear-timescale mass transfer). The mass-radius 
exponent in the third panel is defined as $\zeta_{\rm th} = d \rm{log} R_{\rm th;eq} / d \rm{log} M_{\rm don}$ 
as derives directly from the results in the second panel. 
Horizontal lines show the mass-radius exponent of the donor's Roche lobe $\zeta_{\rm RL} = d \rm{log} R_{\rm RL} / d \rm{log} M_{\rm don}$ assuming a fully non-conservative mass transfer.
The behavior of $\zeta_{\rm th}$ as a function of 
the remaining donor mass can be linked to various features found in binary models, see text.
}
\label{fig:SnS_MT_dzetas}
\end{figure}

In general, the behavior of $R_{\rm th;eq}$ in Fig.~\ref{fig:MT_transition} could be described as a combination of a steep increase, that may happen right after the RLOF (cf. the LMC and Solar cases), followed by a  gradual increase, maximum, and then a gradual decrease of $R_{\rm th;eq}$ as a function of the decreasing donor mass.\footnote{Note that any phase of $R_{\rm th;eq}$ increase will saturate if a model reaches the Hayashi track (at $\sim 1000 \rsun$) and may no longer expand, leading to a $R_{\rm th;eq}$ plateau in Fig.~\ref{fig:MT_transition}.}  The initial steep increase takes $R_{\rm th;eq}$ to the radius that a normal single stellar model would expand to by the end of the HG phase. The SMC donor in Fig.~\ref{fig:MT_transition} was already near that size when RLOF occurred, hence no steep increase in $R_{\rm th;eq}$. The subsequent gradual increase followed by a gradual decrease in $R_{\rm th;eq}$ can be understood as a result of the changing envelope to core mass ratio, as described in Sec.~5.4 and Fig.~9 of \citet{Farrell2022}.

Fig.~\ref{fig:MT_transition} suggests that the equilibrium radius $R_{\rm th;eq}$ is to some extent affected by the 
underlying He abundance profile of the donor \citep[as also found by][]{Quast2019},
in particular at the location of the already mentioned He plateau
(where the $R_{\rm th;eq}$-$M_{\rm don}$ slope of the low-Z models temporarily flattens). 
We illustrate this better in Fig.~\ref{fig:SnS_MT_dzetas}, where we repeat the mass-stripping experiment for the case of a $25 \msun$ 
donor at LMC metallicity and five different initial orbital periods.
The corresponding binary models have led to various outcomes (hot stripped star, partially stripped star, or
nuclear-timescale mass transfer), as explained in the legend.
In the third panel we now show the mass-radius exponent $\zeta_{\rm th} = d \rm{log} R_{\rm th;eq} / d \rm{log} M_{\rm don}$. 
The value of $\zeta_{\rm th}$ is a useful magnifying glass to expose any changes in 
the $R_{\rm th;eq}$-$M_{\rm don}$ slope.
For reference, with horizontal lines we show the mass-radius exponent of the donor's Roche lobe 
$\zeta_{\rm RL} = d \rm{log} R_{\rm RL} / d \rm{log} M_{\rm don}$
for a few different mass ratios and fully non-conservative mass transfer (Eqn.~16.25-26 from \citealt{Tauris2006}).
While the values of $\zeta_{\rm th}$ and $\zeta_{\rm RL}$
can be compared to assess whether the mass transfer will proceed on the thermal timescale (i.e. $\zeta_{\rm th} > \zeta_{\rm RL}$), we stress that this comparison is only valid when the star is in thermal equilibrium, e.g. during a nuclear-timescale expansion or nuclear-timescale mass transfer.

The double-peaked behavior of $\zeta_{\rm th}$ in Fig.~\ref{fig:SnS_MT_dzetas}
can be linked to some of the features found in binary models throughout Sec.~\ref{sec:results}.
In most binaries that evolve through phase(s) of nuclear-timescale 
mass transfer, the donor regains thermal-equilibrium when 
$R_{\rm th;eq}$ drops strongly as a function of $M_{\rm don}$
(between $\sim 14$ and $17.5 \msun$ in Fig.~\ref{fig:SnS_MT_dzetas}).
In turn, in binary models in which the donor remains detached for most of the core-He burning lifetime 
(either as hot- or partially-stripped star) the thermal-timescale mass transfer tends to strip
the donor down to deeper layers of constant $X_{\rm He}$ ($\lesssim 14 \msun$ in Fig.~\ref{fig:SnS_MT_dzetas}).
The abundance pattern of these layers sets 
the surface composition of partially-stripped stars discussed in Sec.~\ref{sec:res_partstrip}.

We always find a $\zeta_{\rm th}$ peak in layers above the He abundance plateau.
Its saw-like shape is related to the step-like He abundance gradient
above the region of He plateau. It manifests itself in oscillations of the mass transfer rate during 
the nuclear-timescale mass transfer phase when a donor is being stripped from those envelope layers 
(see e.g. the SMC models in Fig.~\ref{fig.diffQ}). 
The local minimum of $\zeta_{\rm th}$ around $M_{\rm don} \sim 14 \msun$ in Fig.~\ref{fig:SnS_MT_dzetas}, 
on the other hand, is responsible for the secondary major $\dot{M}$ peak that in some models is a very 
prominent feature (see Fig.~\ref{fig:sns_30example} and Fig.~\ref{fig:sns_30exampleHRD}, label 'D'). 
\footnote{If the mass ratio is close to unity then $\zeta_{\rm th}$ may temporarily become smaller than $\zeta_{\rm RL}$, 
leading to a brief phase of another thermal-timescale mass transfer.}
Stripping through the bottom envelope layers of constant He abundance ($M_{\rm don} \lesssim 14 \msun$)
leads to a smooth $\zeta_{\rm th}$ increase and also results in a smooth $\dot{M}$ behavior 
found in late phases of mass transfer in binary models. \footnote{We notice a slight shift in the values 
$\zeta_{\rm th}$ of those layers between the HG and more evolved core-He burning donors in Fig.~\ref{fig:SnS_MT_dzetas}. 
This is likely associated with changes in the H abundance and He/H gradient 
in the moving location of the H-burning shell, which has been found to affect the radii of core-He burning massive stars \citep{Langer1991}.}

It is noteworthy that the values of $\zeta_{\rm th}$ found in the mid-envelope region
can be very high, with $\zeta_{\rm th} > \zeta_{\rm RL}$ for even extreme mass ratios of $q < 0.1$. We discuss this in the context of ULX sources with NS accretors in Sec.~\ref{sec:disc_ulx}.

In summary, we find that partial envelope stripping and nuclear-timescale mass transfer occurs in binary models 
when the donor can be in thermal equilibrium as a partially-stripped star of an intermediate size
($R_{\rm th;eq}$ of a few tens or hundreds $\rsun$). Such donors, when evolved as single stars, 
are characterized by a relatively modest HG expansion and a long blue/yellow supergiant 
lifetime during the core-He burning phase. The post-MS expansion of a massive star is thus the key factor affecting 
its evolution in an interacting binary. A secondary role is played by detailed features of the He abundance profile, 
some of which may be causing small variations in the mass transfer rate or temporary detachments.
The abundance profile in massive star envelopes is a particularly uncertain prediction of stellar 
models. Consequently, details of the nuclear-timescale mass transfer sequences should be treated with much caution.

\section{Discussion}
\label{sec:discussion}

\subsection{The importance of post-MS expansion and its uncertainty due to mixing}
\label{disc.uncertainty}

Throughout the paper, we have found that the course and outcome of stable mass transfer 
evolution initiated by a post-MS donor is closely related 
to the way in which massive stars expand after the end of MS when they transition 
to the core-He burning phase. 
Stars that rapidly expand all the way to the red (super)giant branch 
(i.e. the HG phase),
when transferring mass in interacting binaries, 
become nearly fully stripped of their H envelopes 
in a short phase of thermal-timescale mass transfer 
($\mdot \sim 10^{-3}\msunyr$, $\Delta T_{\rm MT} \lesssim 10^4 \, \rm yr$). 
In contrast, stars that remain relatively compact in the transition to the core-He burning phase and 
burn He as blue or yellow supergiants, when in binaries, remain only partially stripped of their H envelopes 
(at least by the end of core-He burning) and can evolve through phases 
of nuclear-timescale mass transfer ($\mdot \sim 10^{-5}\msunyr$, $\Delta T_{\rm MT} > 10^5 \, \rm yr$). 
Such a halted HG expansion of massive stars has been found in single low-Z models 
computed with different codes over the years \citep[being more and more prominent 
the lower the metallicity, e.g.][]{Brunish1982,Baraffe1991,Langer1991,Georgy2013,Tang2014,Groh2019,Klencki2020}.
The fact that low-Z massive giants can remain much more compact compared to their high (Solar) metallicity 
counterparts is a result of a complicated interplay between at least two different Z-dependent factors:
higher temperatures and densities of low-Z helium cores at TAMS as well as lower opacities of low-Z envelopes. 
It is not until the current study that this phenomenon has gained special significance as the key factor affecting 
the evolution of stars through mass transfer in binaries.

It is essential to realize that the post-MS expansion of massive stars is notoriously model sensitive 
due to being highly dependent on the efficiency of internal mixing. This well-established fact \citep[e.g.][]{Langer1985,Langer1995,Maeder2001} is especially clear in the recent studies in which the increased computational power has allowed for a more comprehensive exploration of various mixing coefficients  \citep{Schootemeijer2019,Klencki2020,Kaiser2020,Higgins2020}.
Among the factors 
that were identified to play a role are convective-core overshooting during the MS \citep{Stothers1992,Langer1995}, semiconvection
\citep{Langer1985,Langer1991}, rotational mixing \citep{Georgy2013}, as well as past accretion phases 
\citep[especially if the accretor is non-rejuvenated, see][]{Hellings1984,Braun1995,Dray2007}.
We refer to an extensive discussion of the subject in Sec.~5.1 and App.B in \citet{Klencki2020}.
For example, if we were to compute our grids of binary models with lower efficiency of semiconvection 
($\alpha_{\rm SC} = 1$ instead of $\alpha_{\rm SC} = 33$) then even the SMC-metallicity models 
would all experience a rapid HG expansion all the way until the red supergiant stage (and be subject to full
envelope stripping through thermal-timescale mass transfer).

For the time being, the key piece of evidence in support of the models presented in this paper are large populations 
of blue and yellow supergiants identified in the LMC and the SMC (\citealt{Hunter2008,Neugent2010,Neugent2012,Urbaneja2017,Kalari2018}, 
see also HR diagrams in \citealt{Ramachandran2019,Gilkis2021}). Their existence can only be reconciled 
with models that predict a significant fraction of core-He burning taking place in the middle of the HR diagram 
\citep{Schootemeijer2019,Klencki2020}
such as the models adopted here (with a halted HG expansion at the LMC and SMC compositions).
A robust comparison between theory and observation to calibrate the post-MS expansion
is challenged by the fact that the population of blue and yellow supergiants may also include 
stars in a post-red supergiant stage 
\citep[either stripped in binaries or through cool-star winds, or stars experiencing blue-loops, e.g.][]{Ekstrom2012,Meynet2015,Farrell2019}
as well as potentially some of the 
accretors and mergers from past binary interaction phases \citep[e.g.][]{Podsiadlowski1992,Vanbeveren2013,Glebbeek2013,Justham2014}.
Extended grids of binary models, exploring various mixing assumptions and including the evolution of accretors, 
are likely necessary to further calibrate the post-MS expansion of LMC and SMC stars in the future studies. 

Instead, a promising way to constrain the mass transfer evolution in low-Z binaries and verify our findings 
is to search for signatures of partial-envelope stripping and nuclear-timescale mass transfer among the populations 
of massive stars in metal-poor galaxies. We outline the main observational predictions from our models in 
Sec.~\ref{sec:disc_CHEB_HRD}.

Notably, while details of the post-MS expansion of massive stars at a given metallicity are very uncertain 
in stellar models, the overall trend with metallicity appears to be a robust prediction \citep[Sec.~5.2 in][]{Klencki2020}.
As such, we predict that at some sufficiently low metallicity (possibly already at the LMC composition), 
the thermal-timescale mass transfer and full envelope stripping will yield ground to nuclear-timescale mass exchange 
and partial-envelope stripping of massive post-MS donors.  

\subsection{The case of stellar-accretor binaries}
\label{sec.applicability_stellar_binaries}

In this work, for simplicity, we treated the secondary star as a point mass and we set the accretion efficiency $\beta$
at the Eddington limit. This makes our models directly applicable to BH binaries (assuming that the Eddington limit is not 
substantially exceeded). However, as we argue below, we predict that most of our results and conclusions will also hold 
for the case of binaries with stellar accretors. 

It is important to realize that Eddington-limited accretion in massive binaries is similar to the assumption of 
a fully non-conservative mass transfer (i.e. $\beta = 0$). 
For the default initial mass ratio $q = 0.6$, the accretor masses range from 
$6$ to $31.8 \msun$ across our models. The corresponding Eddington accretion rates for BHs (assumed as the accretion limit in our models) range from $\sim 1.3 \times 10^{-7} \msunyr$
to $\sim 7 \times 10^{-7} \msunyr$. Even during phases of slow nuclear-timescale mass transfer, with typical rates of $10^{-5} \msunyr$, 
this yields accretion efficiencies of only a few percent. We thus expect that a grid of binary models with stellar 
accretors and a small accretion efficiency would produce mass-transfer sequences very similar to our models. 

As we discuss in Sec.~\ref{disc.accretion_eff}, 
the true accretion efficiency in stellar-accretor binaries corresponding 
to our models is currently unknown and so ideally any $\beta$ value 
between $0$ and $1$ should be considered. Varying $\beta$, and similarly considering different values for 
the specific angular momentum of the non-accreted matter, would affect the evolution of binary separation 
during mass transfer and consequently the size of the donor's Roche lobe. 
These effects are degenerate with changing the mass ratio between the binary components. 
In Fig.~\ref{fig.diffQ}, using the $25 \msun$ donor example,
we showed that for a wide range of initial mass ratios ($q$ between $0.25$ and $1.5$), 
the essential differences between the Solar and low-Z models remain unaffected. 
In particular, binaries in which thermal-timescale mass transfer produces fully stripped donors (the Solar-Z example)
evolve towards the same outcome no matter the mass ratio. 
In the case of low-Z models with partial envelope stripping, variations in factors that affect the Roche lobe 
of the donor will likely affect the ratio between systems that remain primarily in the detached stage 
and those that maintain the nuclear-timescale mass transfer. 

Besides the possibility of high accretion efficiencies, the presence of a stellar accretor could also lead 
to some of the binaries evolving towards a contact phase \citep{Pols1994,Wellstein2001} and potentially dynamical 
instability and a merger \citep{Marchant2016}. Details of this process remain highly uncertain. 

\subsection{Increased accretion efficiency in low-Z binaries?}
\label{disc.accretion_eff}

It is interesting to speculate about the impact of long and slow mass transfer phases in our low-Z models 
on the fraction of the transferred matter that gets accreted by the companion.
A clear coherent picture of accretion efficiencies $\beta$ in binaries of different orbital periods and component masses 
is still missing. Observational clues from double-lined eclipsing SMC binaries \citep{deMink2009}, WR-O star systems
\citep{Petrovic2005}, the well studied sdO+Be binary system {\ensuremath{\varphi}} Persei \citep{Schootemeijer2018_persei}, 
or Be-X ray binaries in the SMC \citep{Vinciguerra2020} yield  very different accretion efficiencies, ranging 
from highly non-conservative ($\beta \approx 0$) to nearly fully conservative cases ($\beta \approx 1$).\footnote{Note that these different 
types of post-interaction systems are probing different mass transfer regimes, so 
the large spread in $\beta$ values is perhaps not unexpected.}

Since the emergence of binary evolution models with rotation \citep{Langer2003}, a promising 
way of obtaining $\beta$ self-consistently from models has been to assume that the material can be accreted 
conservatively up to the point when the accretor becomes spun-up to critical rotation, at which 
point $\beta$ drastically decreases and is further controlled by the timescale of internal angular momentum 
transport in the accretor \citep{Langer2003_IAUS}.\footnote{A model of efficient accretion 
at breakup rotation velocity has been proposed by \citep{Popham1991}.}
Because little mass accretion is needed to reach the critical rotation of the accretor \citep{Packet1981}, 
and the thermal timescale of mass transfer is typically much shorter than the timescale for angular momentum 
transport in the accretor, such models tend to predict a highly non-conservative post-MS mass transfer evolution 
\citep{Petrovic2005,deMink2013}. The exception are cases with a close to equal mass ratio due 
to their relatively lower mass transfer rates \citep[e.g.][]{Cantiello2007}.

Based on those arguments, we expect that long nuclear-timescale mass transfer phases in low-Z binaries 
may result in considerably higher accretion efficiencies compared to systems evolving through 
only thermal-timescale mass transfer. This might possibly bring such models into agreement with $\beta \sim 0.5$ inferred 
from Be-X-ray binaries of the SMC composition \citep{Vinciguerra2020}, although a detailed study of the issue
is certainty needed before any conclusions could be drawn. Longer mass transfer phases 
could potentially also extend the duration of the Be phenomenon in rapidly-rotating accretors, which might 
help in resolving the apparent overabundance of Be stars in the SMC compared to the Galactic environment \citep{Dray2006}.

\subsection{Binary-interaction products in the HR diagram}

\label{sec:disc_CHEB_HRD}

\begin{figure*}
\centering
\includegraphics[width=0.9\textwidth]{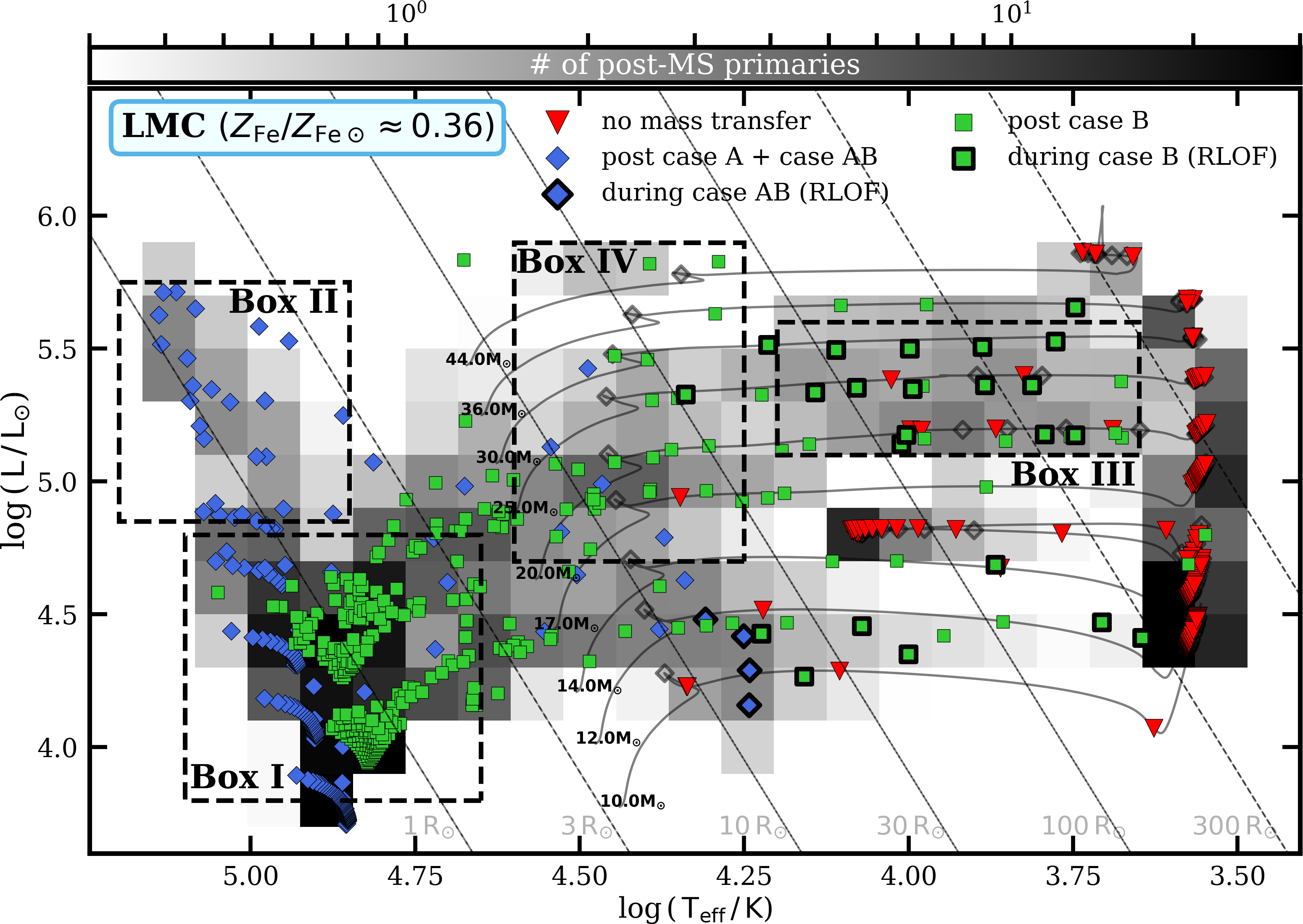}	
\caption{Distribution of post-MS primaries in the HR diagram (the vast majority being at the core-He burning stage) inferred from the grid of $q = 0.6$ binary models 
at the LMC composition. For the SMC version of this figure see \ref{fig:disc_HRD_hist2d_SMC}. 
Models have been weighted by the initial mass function 
and orbital period distribution of early-type binaries, and normalized to a constant star formation rate 
of 1 $M_{\odot}/\rm yr$ (see text). 
The overall distribution is showed in greyscale. In addition, we plot a random sampling of stars from the distribution, color-coded according to their pre- or post-interaction state. 
Several single stellar tracks are plotted with solid lines for comparison. Around $30\%$ of primaries reside in wide non-interacting systems. 
The four Boxes indicate four characteristic regions: hot stripped stars from both case A + case AB evolution as well as those formed through 
case B mass transfer (Box I), hot stripped stars from case A + case AB mass transfer only (Box II), a region where 
most of the stars are donors in currently mass-transferring systems (Box III), and finally a region where most of the long-lived partially-stripped stars 
in detached binaries reside (Box IV).
}
\label{fig:disc_HRD_hist2d}
\end{figure*}

Here, we discuss one of the most interesting implications of binary evolution models: the predicted 
distribution of post-MS stars, including the binary-interaction products, in the HR diagram. 
In the current study, for each initial composition we computed only a single grid of models (with initial mass ratio 
$q = 0.6$) and evolved only the primary stars. For that reason, we are unable to construct a complete population model 
of massive stars and binaries. Even a single $q$ grid, however, is sufficient to showcase the main subpopulations 
of post-MS primaries predicted by the binary models and discuss their main observational characteristics. 
To this end, we choose the LMC grid as the most illustrative example. 

In order to construct a simple population model, we weight all our binary models by the initial mass function ${\rm d}N/{\rm d}M_1 \propto M_1^{-2.3}$ 
\citep{Salpeter1955,Bastian2010}, making the usual assumption that it well describes the initial mass function of primaries \citep{Kroupa2013}.
In addition, we weight each model according to an initial orbital period distribution 
${\rm d}N/{\rm d \, log} P_{\rm ini} \propto {\rm log} P_{\rm ini}^{-0.55}$ \citep{Sana2012}.
We normalize the period distribution to the range ${\rm log} P_{\rm ini} = [0.15, 5.5]$.
While wider early-type binaries do exist, they are very rare for $q > 0.3$ \citep[see Fig.~37 in][]{Moe2017}.
For simplicity, we only consider massive stars that are formed in binaries, which neglects that a significant fraction 
of early-type stars are formed in triples or higher order systems \citep[see Fig.~39 in][]{Moe2017}.
For each primary mass, we use our widest binary model (i.e. the non-interacting one) to represent all the wide non-interacting 
binaries in the weighting procedure described above. Likewise, we extrapolate to the shortest orbital periods of very close binaries
($P_{\rm ini} \sim 1.41$ days) using the model with the shortest initial orbital period in our grid ($P_{\rm ini} \approx 3.16$ days), 
which is relevant for our estimates of the products of case A mass transfer evolution. 
We caution that this is a rather crude approximation: very close binaries will produce somewhat less massive and luminous stripped 
helium stars than our $P_{\rm ini} \approx 3.16$ days models. \footnote{Additionally, their orbits would typically be synchronized at ZAMS, 
which could lead to chemically-homogeneous evolution of the most massive LMC or SMC cases in that period range \citep{deMink2009_che}.}

Based on these assumptions, we are able to estimate the distribution of post-MS primaries in the HR diagram.
The result is shown in greyscale in Fig.~\ref{fig:disc_HRD_hist2d}, normalized to show a number of stars per bin assuming a constant star-formation rate of $1 M_{\odot}/\rm yr$.\footnote{The normalization assumes that all massive stars are formed in binaries and that the initial mass ratio is 0.6.} In addition, we plot a random sampling of 770 primaries from the distribution (the expected number, given the normalization) and color-code the stars according to their pre- or post-interaction state.
For comparison, with grey lines we plot several single stellar tracks. 

Post-MS primaries from wide non-interacting binaries (around $30\%$ of stars in Fig.~\ref{fig:disc_HRD_hist2d}) occupy regions in the HR diagram that would also be populated by single stellar tracks, primarily the red (super) giant 
branch at $\logteff \approx 3.6$. Binary-interaction products, on the other hand, can occupy nearly any location in the HR diagram
(although in the region of the MS they would be a small minority compared to core-H burning objects).
Below we describe four most prominent categories of post-interaction and interacting primaries, with Boxes I-IV 
marking regions where they can typically be found in Fig.~\ref{fig:disc_HRD_hist2d}.

Box I is populated by hot fully stripped stars originating from both the primaries that interacted already
during the MS (case A mass transfer) as well as those 
that are stripped solely through post-MS (case B) mass transfer. As the case A mass transfer systems are not the focus of the paper and have 
not been presented as part of the results,
we clarify that 
in each of our case A models the primary becomes a fully stripped helium star after a sequence of case A and case AB mass transfer 
phases. This is in agreement with all the previous detailed mass transfer models of such systems 
\citep[e.g.][]{Podsiadlowski1992,Pols1994,Wellstein2001,Petrovic2005,deMink2007,Wang2020}. In general, we find 
that stars stripped in case A + case AB mass transfer are slightly hotter in our models compared to those fully stripped in 
post-MS mass transfer (by about $0.1$ dex in $\logteff$). A more detailed discussion of differences between the two classes, 
including potential surface abundance signatures \citep[e.g.][]{Schootemeijer2018}, is beyond the scope of this paper. 
Notably, Box I is populated by primaries with initial masses $\leq 14 \msun$ from our LMC grid, i.e. the mass range 
in which post-MS mass transfer always leads to full envelope stripping on the thermal timescale (see Fig.~\ref{fig:tricolor} 
for the summary of model outcomes).

In contrast to Box I, the hot stripped stars populating Box II (WR stars above a certain luminosity) are only formed in case A mass transfer systems 
(to compare, see the right-hand panel of Fig.~\ref{fig:HRD_PShist2d} which excludes the case A systems). This is because 
in the $M_1 \geq 17 \msun$ primary mass range the majority of LMC models evolve to either produce long-lived partially stripped 
stars or spend most of the core-He burning lifetime in nuclear-timescale mass transfer. 
This naturally leads to a prediction that the formation efficiency of hot stripped stars in interacting binaries decreases 
with increasing luminosity around $\logL \approx 4.8$ for the LMC composition relative to Solar metallicity models (which always lead 
to full envelope stripping in mass transfer). We discuss this further in Sec.~\ref{sec:disc_strip_form}.

Box III is where the majority of donors in nuclear-timescale mass transferring systems reside. They originate from 
systems with primaries in the mass range where single stellar models predict a halted HG expansion and core-He burning 
blue or yellow supergiants ($M_1$ between $\sim 20$ and $36 \msun$ in our LMC grid). In fact, about 
$40\%$ of stars in Box III are currently mass transferring, i.e. they are donors in semi-detached or contact systems.
Their presence or absence in a population of late-blue/yellow supergiants in the LMC (or similarly SMC) will serve 
as a strong observational test to verify whether nuclear-timescale post-MS evolution can happen already 
at the LMC (or SMC) metallicity. Currently, the census of yellow supergiants in the Magellanic Clouds comes from single-epoch spectroscopic surveys by \citet{Neugent2010,Neugent2012} and offers little insights into their binary fraction.
The period distribution of mass-transferring systems in Box III 
is roughly flat in log($P$) in the range between $\sim 50$ and $1000$ days. The long periods of these systems
may make it challenging to assess their binary nature in typical spectroscopic surveys \citep[unless a long-term monitoring campaign is possible, e.g.][]{Sperauskas2014}. An alternative approach could be to systematically search for eclipsing binaries with yellow supergiants in photometric variability surveys. One such system was serendipitously discovered in a dwarf galaxy Holmberg IX by \citet{Prieto2008} with the Large Binocular Telescope. The authors also pointed out to another candidate system in the SMC that had already been present in the All Sky Automated Survey data.

Finally, Box IV is largely populated by the class of long-lived partially-stripped core-He burning stars
in detached binaries that is unique to our low-Z models,
see also Sec.~\ref{sec:res_partstrip} for details about their formation.\footnote{The grey region just below 
Box IV are stripped donors contracting on the thermal-timescale to become hot stripped stars. }
The properties of stars in Box IV are quite interesting. First of all, 
they are under-massive for their luminosity when compared to single stellar tracks (by about a factor of 2,
depending on luminosity, see Fig.~\ref{fig:PS_masses_LMC}). They are also helium and nitrogen enriched 
while being slow rotators (Fig.~\ref{fig:PS_histograms_LMC}), which distinguishes them 
from fast rotators in which the surface enrichment is the result of rotational mixing 
in their interiors. These properties in principle make them similar to MS donors interacting in case A mass transfer 
binaries. However, in case A systems the donor spends most of its lifetime in a semi-detached or nearly Roche-lobe filling stage 
in a close orbit \citep{Wang2020}. Partially-stripped stars from the post-MS mass transfer models, on the other hand,
can be found in binaries with much wider orbits, with their period distribution stretching from tens to 
thousands of days (Fig.~\ref{fig:PS_histograms_LMC}). Together with the supposed presence of a rapidly-spinning companion (spun-up as a result of mass transfer), 
these set of characteristics makes them a unique class of objects.

It is tempting to view partially-stripped stars as a promising explanation for at least some of the 
puzzling nitrogen-enriched slowly-rotating B- and O-type (super)giants that have been 
identified in VLT-Flames spectroscopic surveys of the LMC \citep{Hunter2008,McEvoy2015,Grin2017}.
Even more interestingly, Box IV stretches out to temperatures cooler than the TAMS, i.e. $\logteff < 4.35$. This 
is the region of the HR diagram that remains virtually unpopulated by single stellar tracks. In reality, a surprisingly 
large number of B (super)giants living apparently next to the MS have been found in the LMC \citep[e.g.][]{Evans2006}, 
many of which being N-rich and with spectroscopic masses systematically lower than their evolutionary masses \citep{McEvoy2015}, 
i.e. with properties resembling those of partially-stripped stars. In the past, binary-interaction channels 
involving the accretors or stellar mergers have been put forth as a possible explanation of this population \citep{Brott2011b,Glebbeek2013}.

The elephant in the room of the discussion so far is the nature of the companion star. Mass losers that become hot stripped stars (Box I and Box II)
emit most of the light in the UV, which is why irrespective of the relative bolometric luminosity of the components,
the secondary (if a star and not a compact object) is nearly always the brighter source in optical bands \citep{Gotberg2018}. 
This is no longer the case for the much cooler partially-stripped stars with $T_{\rm eff} < T_{\rm eff;ZAMS}$ produced 
in our LMC and SMC binary models (e.g. Box III and Box IV). 
Among those, depending on the initial mass ratio and the accretion efficiency, 
there could be systems in which the secondary becomes more massive and luminous than the primary
as well as such in which the primary remains the brighter and more easily detectable component.
In the first case, the partially-stripped companion may remain hidden in the presence 
of a (most likely) rapidly spinning companion with broad emission lines that make it challenging to detect orbital velocity 
variations. Such two components may be especially difficult to disentangle from their spectra in the case of the Box IV
population, where both stars are likely to have a similar spectral type (type B). 
They may appear as more massive analogues of the recently reported LB-1 system \citep{Liu2019_lb1}, 
in which case the presence of a B- and Be-type components in the spectra has proven to be misleading and challenging to unravel 
without a dedicated technique \citep{Shenar2020_lb1,Bodensteiner2020_lb1}.
On the other hand, binaries in which the partially-stripped donor remains the more luminous of the two components
are the most promising to explain some of the peculiar nitrogen-enriched (super)giants mentioned above. 

\subsection{Fewer stripped and WR stars from interacting binaries at low-Z?}
\label{sec:disc_strip_form}

Mass transfer interaction in binary systems is thought to be the main formation channel for stripped stars and, 
alongside the single-star channel in which the envelope is stripped through strong winds \citep{Conti1975,Smith2014}, one of the two main channels 
for their most luminous subclass: WR stars \citep{Paczynski1967}
\footnote{WR stars are essentially stripped stars with high enough luminosity to mass ratio to launch strong optically-thick winds \citep[e.g.][]{Grafener2011}.}.  
While the single-star WR channel is expected to become less efficient the lower the metallicity 
\citep[on the basis that line-driven mass loss decreases with metallicity, see][]{Vink2001,Vink2005}, 
the mass-transfer channel has so far been predicted to have an efficiency that is roughly independent of metallicity
\citep[e.g.][]{Maeder1994}.

\begin{figure}
\includegraphics[width=\columnwidth]{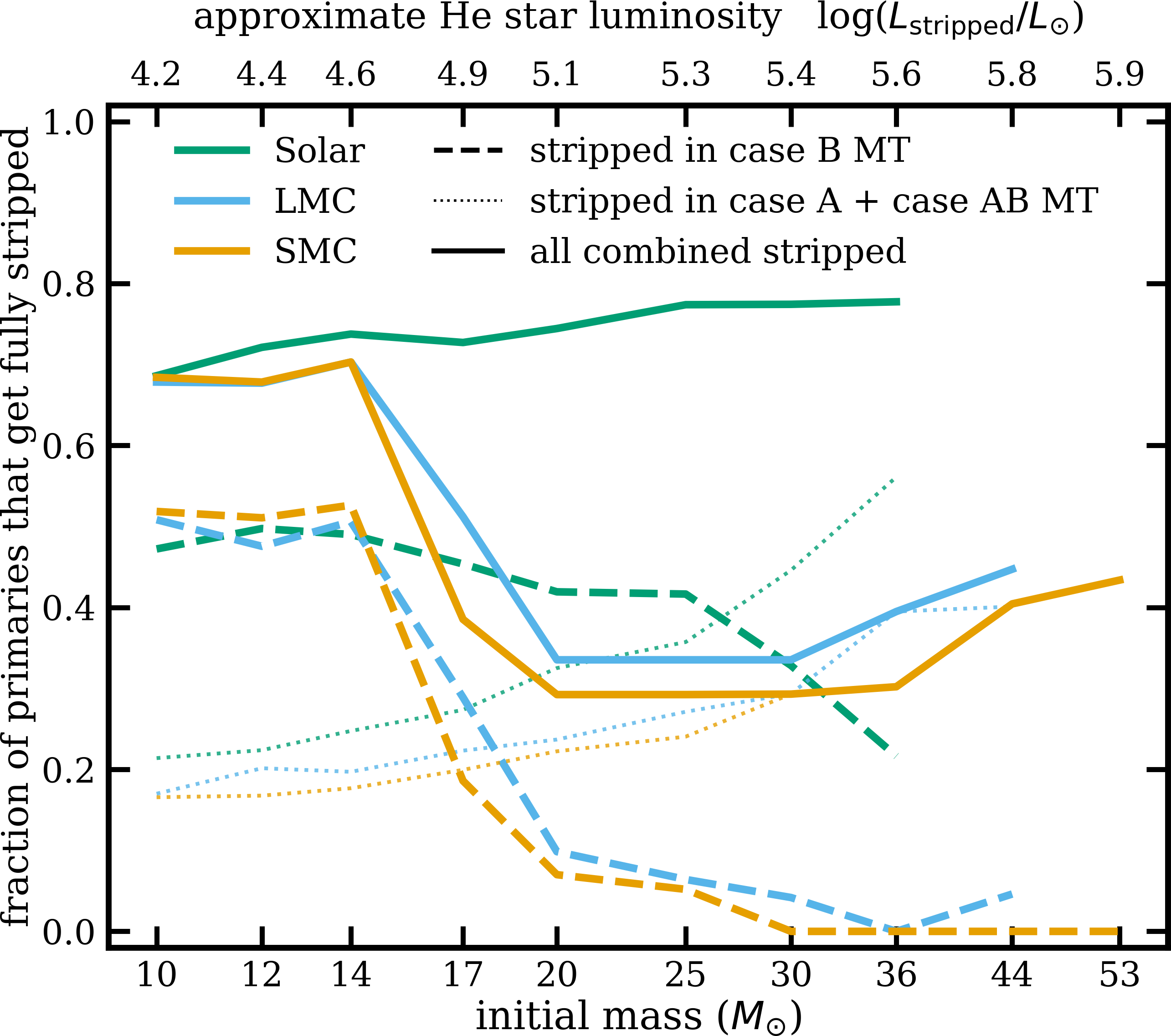}	
\caption{Fraction of primaries that become hot stripped stars as a result of a mass transfer interaction, estimated based on the results from binary 
models as well as the assumed distributions for initial masses and orbital periods (see text). At Solar metallicity, any primary in an interacting binary eventually 
becomes a hot stripped star (the remaining $20\%$-$30\%$ are in wide non-interacting systems). At LMC and SMC metallicities, post-MS donors 
with masses above $\gtrsim 17-20 \msun$ become only partially rather than fully stripped in mass transfer, leading to a predicted decrease in the efficiency 
of formation of stripped stars from low-Z interacting binaries.
}
\label{fig:disc_stripping_vs_mass}
\end{figure}

The binary models computed in this work suggest something different. At LMC and SMC metallicities, most of our post-MS donors with masses above $17 \msun$
never become fully stripped in mass transfer during the core-He burning phase (see Fig.~\ref{fig:tricolor} with the summary of model 
outcomes). As such, we predict the efficiency of the binary formation channel for stripped and WR stars to decrease at low metallicity. 
We illustrate this in Fig.~\ref{fig:disc_stripping_vs_mass}, where for each metallicity, we estimate what fraction of primaries of a given 
initial mass will evolve to form a hot stripped star as a result of a mass-transfer interaction. The rough approximate luminosity of the resulting stripped stars (top X axis) is estimated as the TAMS luminosity of the corresponding single star model.
While at Solar metallicity this fraction is nearly independent 
of the primary mass and very high ($70\%-80\%$, with the remaining $20\%-30\%$ being primaries in non-interacting wide systems), 
at lower LMC and SMC metallicities it drops down by a factor of two (to $30\%-40\%$) for stars with initial masses $\gtrsim 20 \msun$, 
which corresponds to stripped-star luminosities $\logL \gtrsim 4.8$.
For this calculation, similarly to that in Sec.~\ref{sec:disc_CHEB_HRD}, 
we assumed an initial mass function ${\rm d}N/{\rm d}M_1 \propto M_1^{-2.3}$ and an initial orbital period distribution 
${\rm d}N/{\rm d \, log} P_{\rm ini} \propto {\rm log} P_{\rm ini}^{-0.55}$ spanning the range ${\rm log} P_{\rm ini} = [0.15, 5.5]$. 
Based on the few case A mass transfer models in our grid as well as previous detailed studies of case A binary evolution \citep[e.g.][]{Pols1994,Petrovic2005,Wang2020}, 
we predict that all the primaries interacting during the MS will become hot stripped stars after an episode of case AB mass transfer. 
Note that the result in Fig.~\ref{fig:disc_stripping_vs_mass} is based on a single grid of binary models with the initial mass ratio $q = 0.6$. 
The relative fraction of case A compared to post-MS (case B) mass transfer systems is subject to uncertainty in the degree of envelope inflation 
of massive MS stars \citep{Sanyal2015,Sanyal2017,Klencki2020}, a phenomenon that occurs in some of the most massive primaries in our grid of models. 

Fig.~\ref{fig:disc_stripping_vs_mass} constitutes a prediction that the number of stripped stars above a certain mass (and luminosity: $\logL \approx 4.8$ 
in the case of our models) should be lower in metal-poor galaxies compared to high-metallicity environments. 
An observational test may not be straightforward. 
Stripped stars that are not luminous enough to appear as WR stars 
\citep[$\logL < 5.25$ in the LMC and $\logL < 5.6$ in the SMC case, following][]{Shenar2020_WR_vs_Z} are difficult to detect 
in optical surveys. The most promising strategy so far has been far-UV spectroscopy targeting Galactic Be stars and it has recently led to a discovery
of ten new subdwarfs (sdO), thus tripling the number of known Be+sdO systems \citep{Wang2021}. Future similar campaigns focused on the LMC and SMC 
could help verify the metallicity trend predicted by our binary models. 

In the WR regime, the binary fraction of WR stars has long been predicted to increase with metallicity \citep[due to the decreasing efficiency 
of the single-star channel, e.g.][]{Maeder1994}. Surprisingly, this does not seem to be the case as the binary fractions of classical WRs
is found to be about $\sim 40\%$ in the Milky Way and Magellanic Clouds alike \citep{Bartzakos2001,vdHucht2001,Foellmi2003a,Foellmi2003b}.
\citet{Shenar2020_WR_vs_Z} has recently pointed out that this tension may be reduced taking into account the metallicity-dependent 
minimum luminosity for the WR phenomenon. Our results, suggesting a metallicity-depended formation of WRs in binaries, may also help to solve this apparent discrepancy.

\subsection{Implications for explodability and SN progenitors}

\label{sec:disc_SN}

\begin{figure*}
\includegraphics[width=\textwidth]{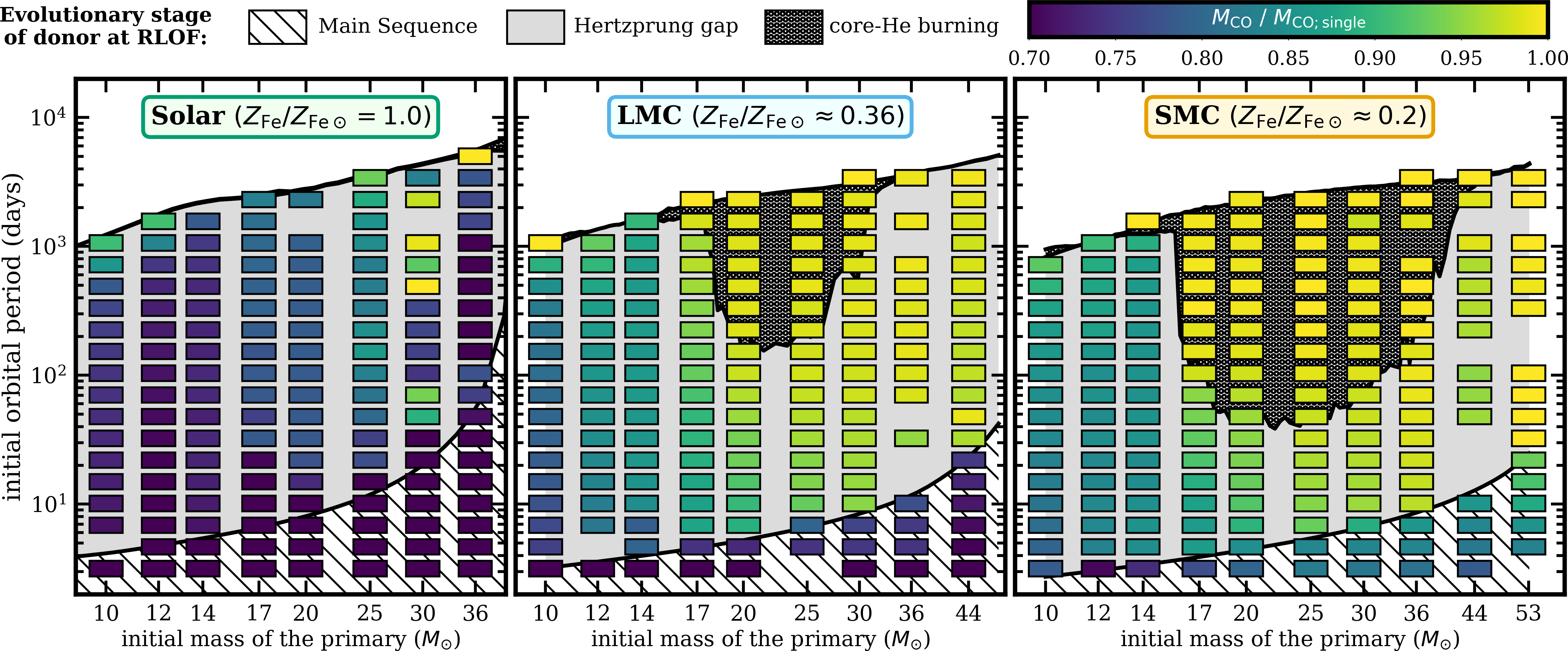}
\caption{Comparison of the CO core masses at the end of core He burning between primaries in binary models ($M_{\rm CO}$) and single stellar models ($M_{\rm CO;single}$). 
\label{fig:square_COcore_mass}
Each binary evolution model, defined by the initial primary mass ($M_1$) and the initial orbital period ($P_{\rm ini}$), is represented by a rectangle colored according to the ratio $M_{\rm CO}/M_{\rm CO;single}$. The initial mass ratio is $q = 0.6$. Single stellar models correspond to $P_{\rm ini} > 10^4$ days. Similarly to Fig.~\ref{fig:square_MTdur}, areas in the background indicate the evolutionary state of the donor stare at the onset of RLOF. We find that the CO core masses in low-Z binary models can be very close to those formed in single stars, despite the mass-transfer interaction. This is because only partial envelope stripping occurs in those models.}
\end{figure*}

In this section, we discuss the impact of the newly found partial envelope stripping of low-Z massive donors on their carbon-oxygen (CO)
cores, pre-SN structure, and explodability.
Whether a massive star loses its H-rich envelope shortly after the end of MS as a result of mass transfer (or strong winds)
or still retains at least part of its envelope during the core-He burning phase has implications for the evolution of the core. 
In the first case, the naked helium core decreases in mass due to wind mass-loss. In the second case, the helium core 
grows in mass due to continuous H-shell burning. This effect was shown to lead to lower CO core masses ($M_{\rm CO}$) and higher carbon mass fractions
($X_{\rm C}$) in the CO cores of fully stripped stars, two factors that largely determine the further core evolution throughout the advanced burning stages 
until the core collapses \citep{Timmes1996,Brown2001,Patton2020}.\footnote{In particular whether or not carbon burning will trigger convection, 
which is especially consequential for the final compactness of the pre-SN core \citep[e.g.][]{Chieffi2020}.}
Consequently, pre-SN stellar models computed from naked helium stars \citep{Woosley1995,McClelland2016,Woosley2019,Ertl2020}
and from fully stripped stars alike \citep{Schneider2021,Laplace2021,Vartanyan2021} were found to have systematically
smaller compactness and to be more prone to explode rather than to collapse into a BH. In particular, \citet{Schneider2021} argues 
that envelope loss due to mass transfer increases the minimum initial mass to form a BH from $\sim$20$-$25$\msun$ up to $\sim70\msun$. 
Among other implications, their work suggests
a drastic reduction of the formation rate of binary BH and BH-NS systems from binary evolution. 
The only exception in \citet{Schneider2021} were cases of case C mass transfer, i.e. the RLOF initiated after the end of core-He burning, 
which the authors found to have very little effect on the pre-SN core structure with respect to single stellar models. 
However, in the case of massive stars with initial masses above 20$-$25$\msun$, case C evolution is
extremely rare except in very metal-poor environments \citep[$Z \lesssim 0.04 \zsun$ in][]{Klencki2020}.

Importantly, the key CO core properties ($M_{\rm CO}$ and $X_{\rm C}$) that were found to determine the compactness
are set already at the end of core-He burning (when we terminate our models). 
This allows us to make a comparison between the single and binary evolution in Fig.~\ref{fig:square_COcore_mass}. Binary models, mapped onto the parameter space of different primary masses and orbital periods, are colored according to the ratio $M_{\rm CO}/M_{\rm CO;single}$ at central He depletion. At Solar metallicity (the left panel), mass transfer interaction and the envelope loss leads to systematically lower CO core masses in primaries compared to single stars of the same initial mass. 
Meanwhile, at the subsolar metallicities of LMC and SMC, we find a large parameter space in which binary models produce CO cores with masses very close to those formed in single stars. Those are the models in which the primaries experience only partial rather than full envelope stripping. Similarly, in Fig.~\ref{fig:square_Xc} we compare the central C mass fractions ($X_{\rm C}$) between single and binary models, finding similarities between partially-stripped primaries and single stars.

Based on Fig.~\ref{fig:square_COcore_mass} and Fig.~\ref{fig:square_Xc} and the results of the above-mentioned studies, it is well-founded to expect that partially-stripped stars 
will produce pre-SN core structures that are similar to those of single stars: more compact and prone to BH formation 
compared to pre-SN cores of fully stripped stars. 
Partial envelope stripping is common among our models: the vast majority of low-Z donors with 
masses between $\sim$20 and 50$\msun$ are never fully stripped in post-MS mass transfer (see Fig.~\ref{fig:tricolor})
and we estimate that fully stripped stars are the minority in this mass range 
(originating mostly from case A mass transfer, see Fig.~\ref{fig:disc_stripping_vs_mass}).
Taken together with the results of \citet{Schneider2021}, this could mean that most of the BHs in 
close binaries in the LMC and SMC-like environments are formed from partially-stripped progenitors.
This may also 'save' the binary BH and BH-NS formation scenario from low-Z massive binaries. 

In the context of SN light curves and stripped SN in general,
we point out that our binary models are terminated at the end of core-He burning and at that point, 
in most of the massive low-Z systems the mass transfer is still ongoing.
This means that even though many of the low-Z primaries have not been fully stripped in our simulations (see Fig.~\ref{fig:square_env_remain} for the final envelope masses), 
they could still lose the remaining hydrogen (or possibly also helium) in the short 
remaining evolution until the core collapse (several thousand yr, depending on mass). Such H-rich layers, 
if not-accreted by the companion, could still reside in the proximity of the star at the core collapse 
and the possible SN, leading to a transient with a circumstellar medium interaction.

\subsection{Implications for ultra-luminous X-ray sources}

\label{sec:disc_ulx}

Here, we discuss the implications of our binary models assuming that the accretor is a compact object: a stellar-mass BH. For the default mass ratio $q = 0.6$, the BH mass would range from $6$ to $\sim32 \msun$ with the corresponding Eddington accretion rates ranging from $\sim 1.3 \times 10^{-7} \msunyr$ to $\sim 7 \times 10^{-7} \msunyr$. This is much lower than the mass transfer rates found in models when they are at the RLOF stage (even during the phases of slow nuclear-timescale mass transfer, when $\dot{M} \sim 10^{-5} \msunyr$). Such BH binaries would thus be HMXBs with highly super-Eddington mass transfer rates. It was shown that in this supercritical regime the thin disk model is no longer valid \citep{Shakura1973}. Instead, the accretion is thought to proceed through a thicker (slim) disk, possibly with a super-Eddington accretion rate and most of the energy being advected into the BH or used to power strong disk winds and jets \citep{Lipunova1999,King2001,Poutanen2007,Lasota2016}. The disk together with optically-thick outflows make the X-ray emission geometrically beamed. Depending on the viewing angle, such systems may appear as ULXs \citep[defined as point X-ray sources with spherically-equivalent X-ray luminosity $L_{\rm X} > 10^{39}$ erg/s, see the review by][]{Kaaret2017} or X-ray bright microquasars such as the Galactic SS433 system \citep{Fabrika2004}.

In this work, we found that a significant fraction of massive low-Z binaries evolve through nuclear-timescale post-MS mass transfer. In such models, the super-Eddington mass exchange phase lasts a few times $10^5$ yr, which is more than an order of magnitude longer than the thermal-timescale mass transfer phase found in the high-metallicity models ($\lesssim 10^4$ yr). This leads to a prediction that HMXBs and ULXs with BH accretors, post-MS donors (blue and yellow supergiants), and periods of at least a few tens of days should be much more common in metal-poor galaxies. 

It is interesting to discuss whether that is indeed observed. At first glace, our results might seem in tension with the fact that no such BH-HMXB system is known in the LMC \citep[the LMC X-1 BH binary contains a MS donor in a short-period orbit][]{Orosz2009}. However, the lack of such systems in the LMC may in fact be statistically consistent with our results. Based on detailed binary models, \citet{Langer2020} estimate that about 100 BH-OB star binaries should be present in the LMC. Taking that the MS lasts for about $90\%$ the stellar lifetime, this corresponds to $\sim 10$ BH binaries with post-MS companions. Roughly $1/3$ of BH binaries predicted by \citet{Langer2020} have companion masses of $20 \msun$ and above and roughly $1/2$ have orbital periods above $\sim50$ days (the minimum donor mass and orbital period for nuclear-timescale mass transfer in our LMC grid). This estimation leads to an approximate prediction of $10/6 \approx 1.7$ BH binaries with post-MS donors that are currently in a mass-transferring state: a number that is not inconsistent with no such systems being observed. 

Instead, the effect of nuclear-timescale mass transfer on the number of ULX sources may be more evident when looking at larger scales, across more distant and diverse galaxies. The recent census of ULX sources in the Local Universe amounts to 629 ULX candidates in 309 galaxies with distance smaller than 40 Mpc \citep{Kovlakas2020}. Interestingly, the number of ULXs per unit of star formation rate is found to increase with decreasing metallicity of the host galaxy (as pointed out by \citealt{Zampieri2009,Mapelli2009}, see also observational studies of the ULX host galaxies by \citealt{Walton11,Swartz11}). Qualitatively, this agrees with the results from out models. Unfortunately, the potential to use the observed number of ULXs to constrain our findings is at the moment hindered by the fact that in the vast majority of ULXs the nature of neither the accretor (whether a BH or a NS) nor the donor (whether or not a high-mass star) is known. 

In the context of ULX systems, it is interesting to highlight that in Sec.~\ref{sec:why_partstrip} and Fig.~\ref{fig:SnS_MT_dzetas} we found that low-Z donors can be characterized by very high $\zeta_{\rm th}$ values in the mid-envelope region (with $\zeta_{\rm th} > \zeta_{\rm RL}$ for even extreme mass ratios of $q < 0.1$).
This 
means that mass transfer from a partially-stripped star could be thermally stable even in binaries 
with a NS accretor ($M_{\rm NS} \approx 1-2.5 \msun$), 
provided that the mass transfer is also dynamically stable, i.e. $\zeta_{\rm ad} > \zeta_{\rm RL}$
(which may well be the case for radiative-envelope supergiants \citealt{Ge2015}).
\citet{Quast2019} suggested this as a possible explanation for ultra-luminous X-ray sources with blue-supergiant 
donors and NS accretors such as the NGC 7793 P13 system \citep{Israel2017}. In their models, the high stability region 
could be reached after stripping most but not quite all of the donor envelope (up to the point with high $\zeta_{\rm th}$
values found near the helium core also in our models). Interestingly, in our stellar models we find another region of high $\zeta_{\rm th}$ located higher up inside the envelope, meaning that a smaller fraction 
of the outer envelope would have to be lost in previous evolution before reaching the high-stability regime. A possible agent 
for such prior stripping in systems with extreme mass radios could be common-envelope evolution with a partial-envelope ejection
(see discussion in Sec.~4.6 in \citealt{Klencki2021}).

\section{Summary}

\label{sec:summary}

In this paper, we studied the mass transfer evolution in massive binaries and the effect played 
by the metallicity of the donor star, as motivated by \citet{Klencki2020}. To this end, using the MESA stellar evolution code, 
we computed grids of detailed binary models at three different metallicities (Solar, LMC, and SMC compositions)
spanning a wide range of orbital periods (from $\sim 3$ to $5000$ days) and initial primary masses 
(from $10 \msun$ to $36$-$53 \msun$, depending on metallicity). Our main focus was on the mass transfer 
initiated by a post-MS donor star.
Due to the challenging numerical nature of such models, we treated the secondary as a point mass. 
Most of the models were computed with an initial mass ratio of $q = 0.6$
although we also explored the effect of varying $q$ for a few cases.
Our conclusions can be summarized as follows.

$-$ We reveal that metallicity has a substantial effect on the course and outcome of mass transfer evolution of massive binaries. 
   While at high (Solar) metallicity
   a post-MS mass transfer is always a short-lived phase ($\Delta T_{\rm MT} \lesssim 10^4 \, \rm yr$)
   of thermal-timescale mass transfer ($\mdot \sim 10^{-3}\msunyr$) with the mass loser becoming a stripped helium star
   \citep[in agreement with the long-standing paradigm, e.g.][]{Paczynski1971,vdHeuvel1975,Podsiadlowski1992,Vanbeveren1998}, 
   this turns out to not be the case in our LMC and SMC models with donor masses $\gtrsim 17\msun$.
   For such massive low-Z donors, the post-MS mass transfer 
   is much less violent: leading either to evolution through long nuclear-timescale mass exchange, 
   that continues until the end of core-He burning ($\Delta T_{\rm MT} \gtrsim 10^5 \, \rm yr$, $\mdot \sim 10^{-5}\msunyr$),
   or to detached binaries with mass-losers that are only partially stripped of their envelopes. 
   
$-$ The origin of the metallicity effect found in the mass transfer models lies in the different response of 
  low-Z donors to mass loss and their small equilibrium radii as partially-stripped core-He burning stars. 
  This in turn is related to the post-MS expansion of massive stars when they transition to the core-He burning stage.
  Stars in which the rapid HG expansion continues until the red (super)giant branch (a common feature 
  of high-Z models), when donors in binaries,  become fully stripped 
  in thermal-timescale mass transfer. Partial envelope stripping and slow mass transfer, on the other hand, 
  occur in the mass range in which massive stars begin to burn He already as blue or yellow supergiants. 
  Such a halted HG expansion is often found in low-Z models of massive stars. Although uncertain due to its sensitivity to mixing (in particular semiconvection), 
  it is supported observationally by the large populations of yellow supergiants in the LMC and the SMC. 
  
$-$ Based on a simple population model, we predict that at SMC and LMC metallicities, fewer (by a factor of $\sim$2$-$2.5) 
  stripped (WR) stars with $4.8 < \logL < 6.0$ are produced by binary interactions compared 
  to a Solar-metallicity environments. This is because among our LMC and SMC models, 
  only $\sim$0$-$20$\%$ of post-MS donors with $M_1 \gtrsim 20 \msun$ become 
  hot stripped stars by the end of core-He burning ($\sim$100$\%$ at Solar metallicity). 
  Case A mass transfer evolution leads to full envelope stripping irrespective of metallicity. 

$-$ We find a significantly longer average duration of post-MS mass transfer in low-Z binary systems 
  (by more than an order of magnitude). In the case of BH accretors, this implies longer lifetimes of high-mass X-ray binaries, which at face value agrees with the large numbers of ULXs found in metal-poor galaxies
  (although the nature of ULX accretors is usually unknown). This also suggests that the immediate progenitors 
  of binary BH systems could be in the mass-transferring state rather than being detached BH-WRs. 
  In the case of stellar accretors, our models 
  provide a testable prediction that many of the blue and yellow supergiants with $\logL \gtrsim 5$ 
  in the LMC and SMC should be in semi-detached binaries ($\sim$30$-$40$\%$ from our $q = 0.6$ grid).
  We also speculate that lower mass transfer rates of low-Z models could lead to higher accretion efficiencies.

$-$ We predict a population of partially-stripped stars in detached binaries in the LMC and SMC. Unlike stripped stars, 
  such mass losers are relatively cool (typically $4.1 < \logteff < 4.5$ and $2.0 < \logg < 4.3$) and thus overlap with the MS and blue supergiants. 
  They are expected to be undermassive for their luminosity (by a factor of $\sim$1.5$-$2), He and N rich, slowly rotating, 
  and reside in binaries with a wide range of periods (from tens to thousands of days). Potentially, they could explain some of 
  the puzzling N-rich slow rotators observed in the LMC \citep{Hunter2008,McEvoy2015,Grin2017}.

$-$ Guided by the carbon-oxygen core properties, we expect the pre-SN core structure of most of the low-Z $\gtrsim 20 \msun$ 
  primaries to be similar to that of single stars in terms of their higher compactness (and lower explodability) 
  compared to fully stripped binary interaction products. 
  Our results therefore suggest that the recently obtained high initial masses needed for the BH formation 
  in case A and case B binaries \citep[$\gtrsim 70 \msun$][]{Schneider2021}, at low metallicity, could potentially be reduced 
  to the often quoted $\sim25\msun$ BH-formation threshold.

\begin{acknowledgements}
We thank the referee for taking the time and effort to carefully review our work.
It is a pleasure to acknowledge 
valuable discussions and suggestions from   Stephen Justham, Selma de Mink, Manos Zapartas, Lida Oskinova, Tomer Shenar, Julia Bodensteiner, Pablo Marchant, Ylva G\"{o}tberg, Thomas Tauris, and David Aguilera-Dena. The authors acknowledge support from the Netherlands Organisation for Scientific Research (NWO). JK acknowledges support from an ESO Fellowship.
\end{acknowledgements}
\bibliographystyle{aa}
\bibliography{ULX_bib.bib}
\begin{appendix}
\section{Additional figures}
\label{app:additional_figures} 

\begin{figure}
\includegraphics[width=\columnwidth]{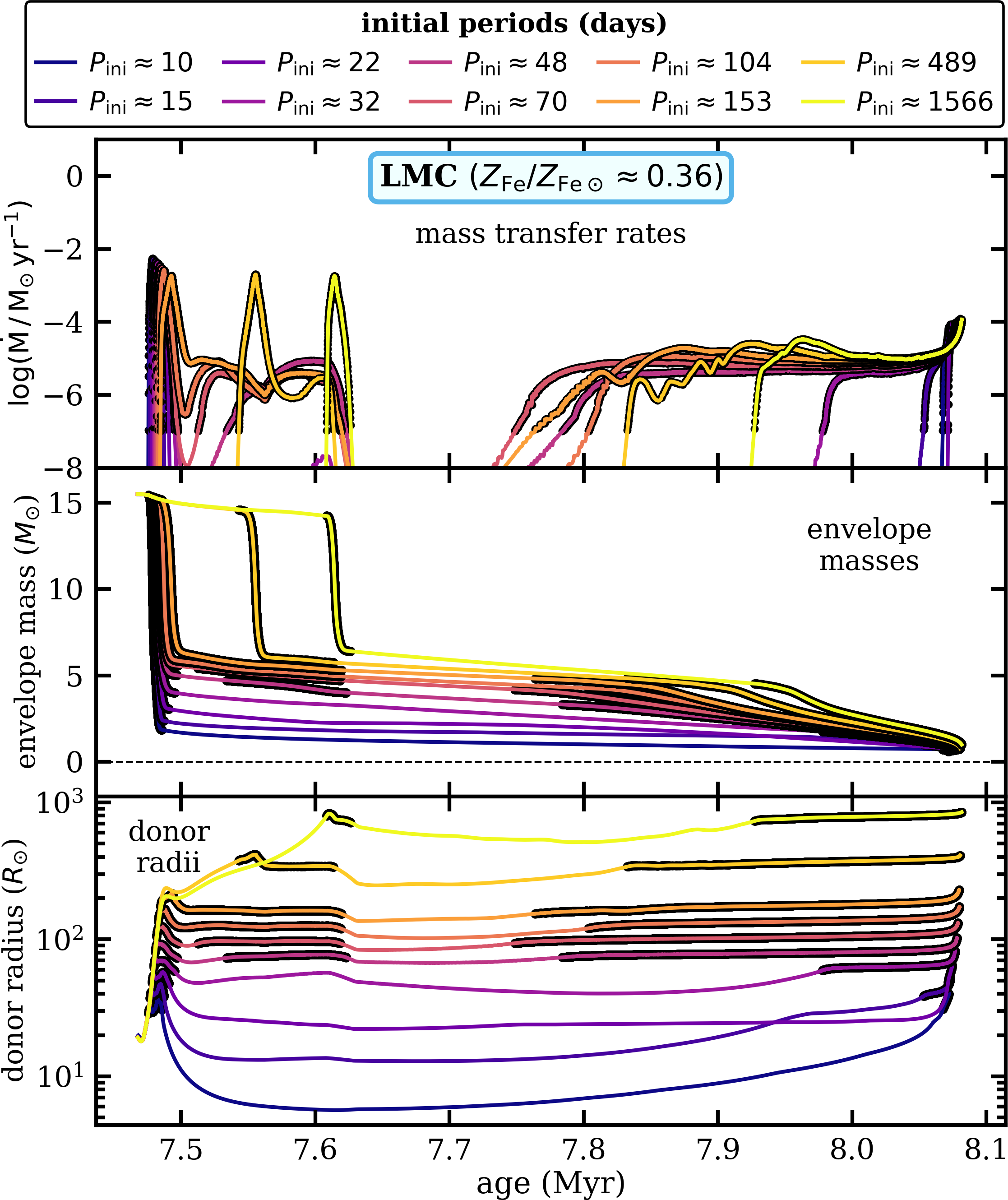}
\caption{Same as Fig.~\ref{fig:period_impact_SMC} but for the LMC metallicity. Mass transfer evolution in binaries with a $25 \msun$ primary (initial mass),  LMC composition, and various initial 
orbital periods $P_{\rm ini}$.}
\label{fig:period_impact_LMC}
\end{figure}

\begin{figure}
\includegraphics[width=\columnwidth]{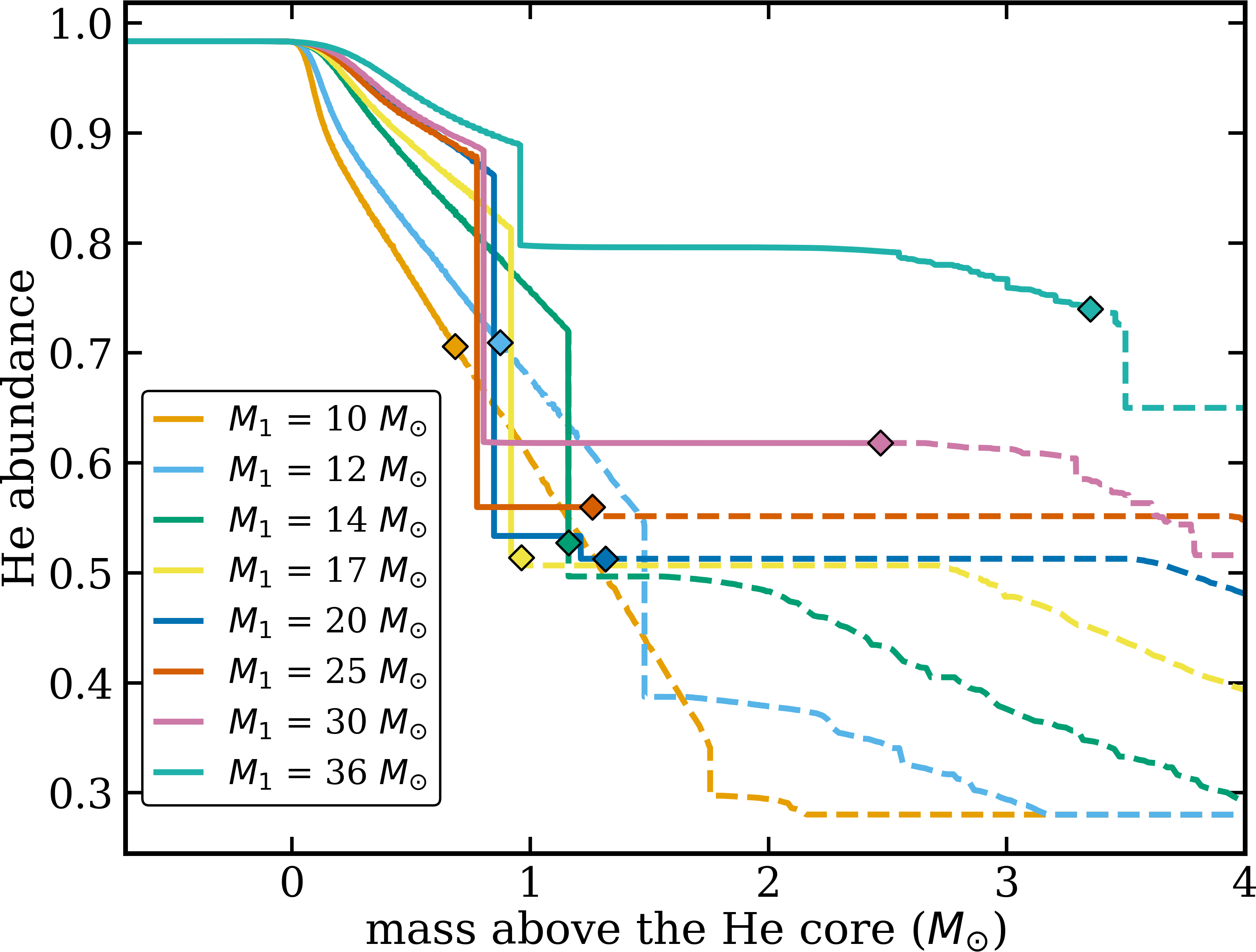}
\caption{Helium abundance profiles near the core-envelope boundary of stripped primaries from a few selected binary models at Solar metallicity. The selected binary models are the same as the ones shown in the left panel of  Fig.~\ref{fig:HRD_PShist2d}. 
Diamonds mark the point to which each donor was stripped in mass transfer. Solid lines are taken from donor structures just after the end of mass transfer. Dashed lines show the extension of He profiles taken from the structure from the onset of mass transfer. The figure illustrates that the more massive the primary ($M_1$), the more massive the envelope that remains after mass transfer. As a result, stripped stars from models with $M_1 \gtrsim 17 \msun$ primaries may retain some envelope layers from the plateau of He abundance, leading to cooler effective temperatures (see Fig.~\ref{fig:HRD_PShist2d} and the associated text).} 
\label{fig:abundance_profiles}
\end{figure}

\begin{figure*}
\includegraphics[width=\textwidth]{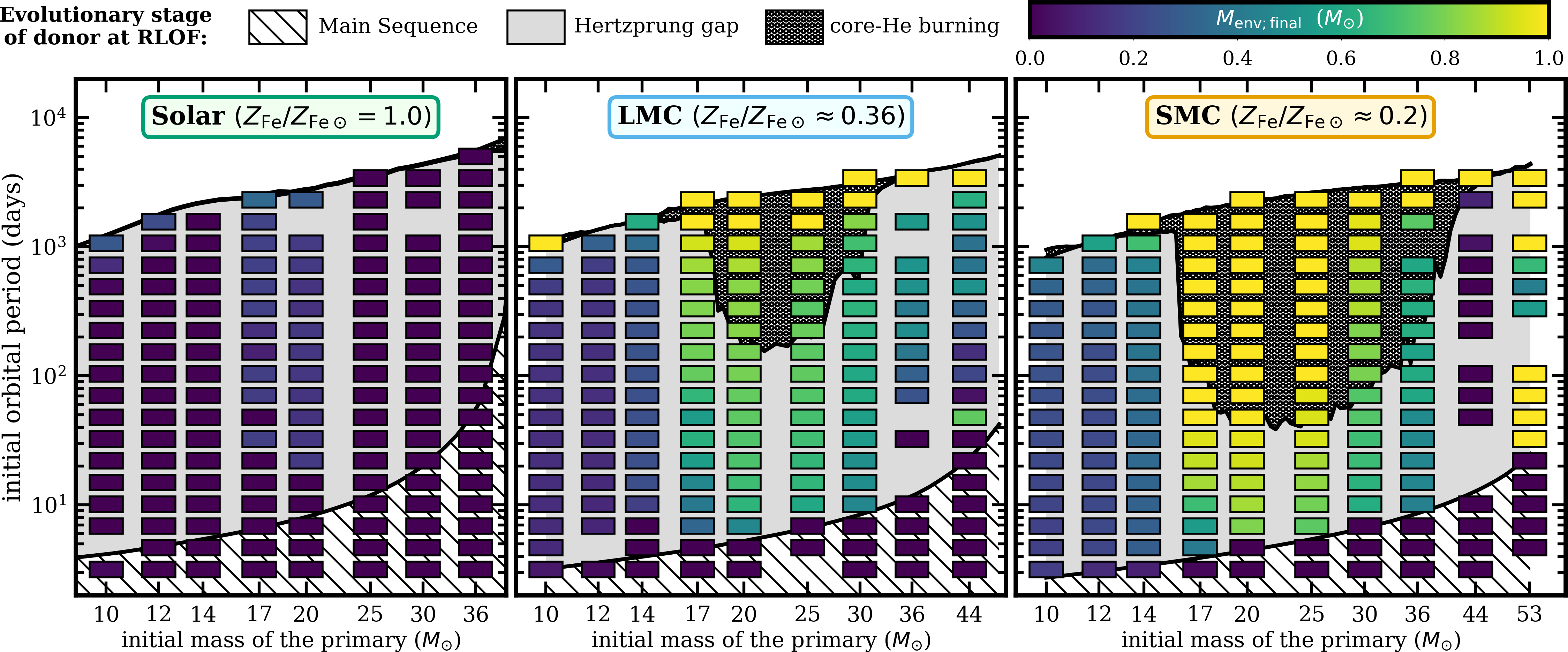}
\caption{Similar to
Fig.~\ref{fig:square_MTdur} but the rectangles representing the binary models are colored according to the mass of the H-rich envelope (H content $X_H > 10^{-3}$) that remains at the end of core-He burning. The colorscale focuses on the range between 0 and $1\msun$ but in some of the low-Z binary models the remaining envelope mass is even greater than $1\msun$.
}
\label{fig:square_env_remain}
\end{figure*}

\begin{figure*}
\includegraphics[width=\textwidth]{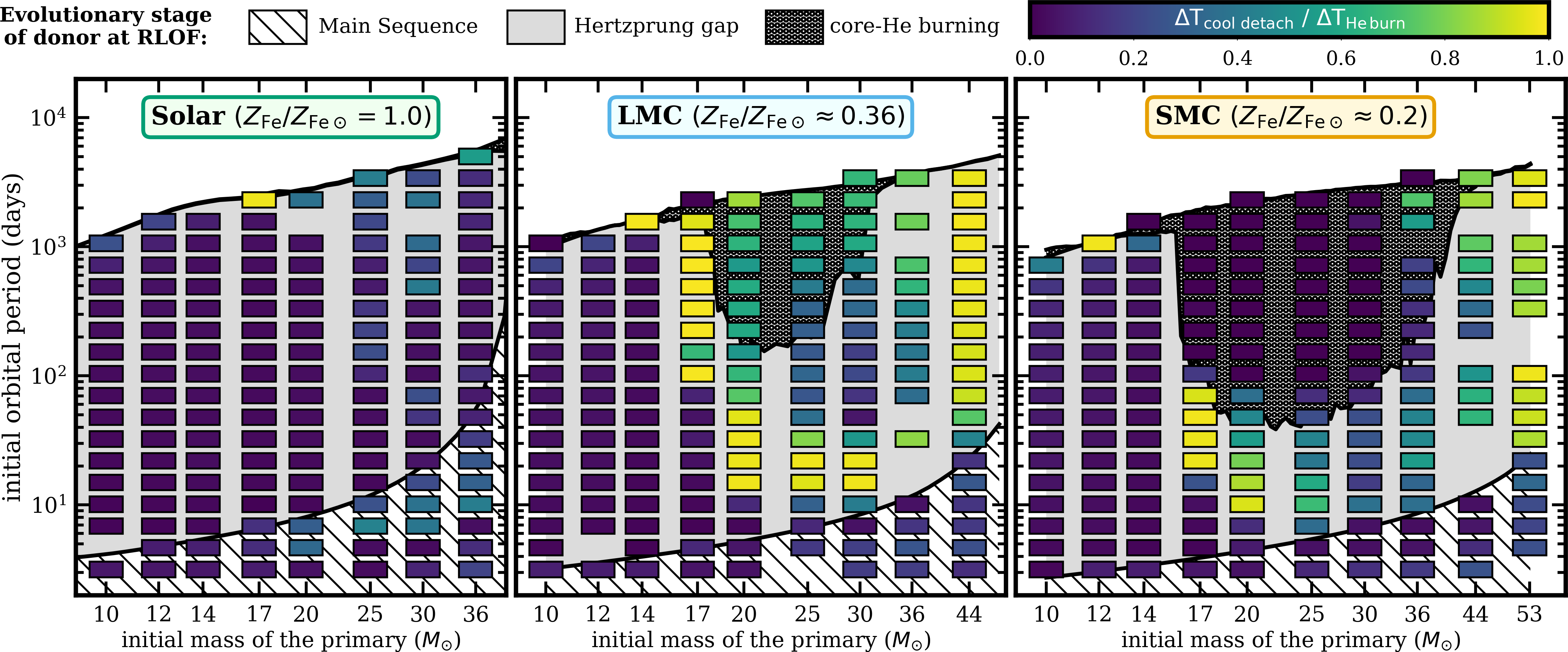}
\caption{Similar to
Fig.~\ref{fig:square_MTdur} but the rectangles representing the binary models are colored according to the integrated duration of a post-interaction detachment phase during which the core-He burning mass loser remains on the cooler side of the ZAMS line in the HR diagram. Rectangles corresponding to models with post-MS donors (either at the Hertzprung gap or the core-He burning stage, see the background colors) that are colored in various shades of green and yellow are models that produce partially-stripped long-lived stars (see Sec.~\ref{sec:res_partstrip} and Sec.~\ref{sec:disc_CHEB_HRD} for a discussion of this population).}
\label{fig:cooldetach}
\end{figure*}

\begin{figure*}
\includegraphics[width=\textwidth]{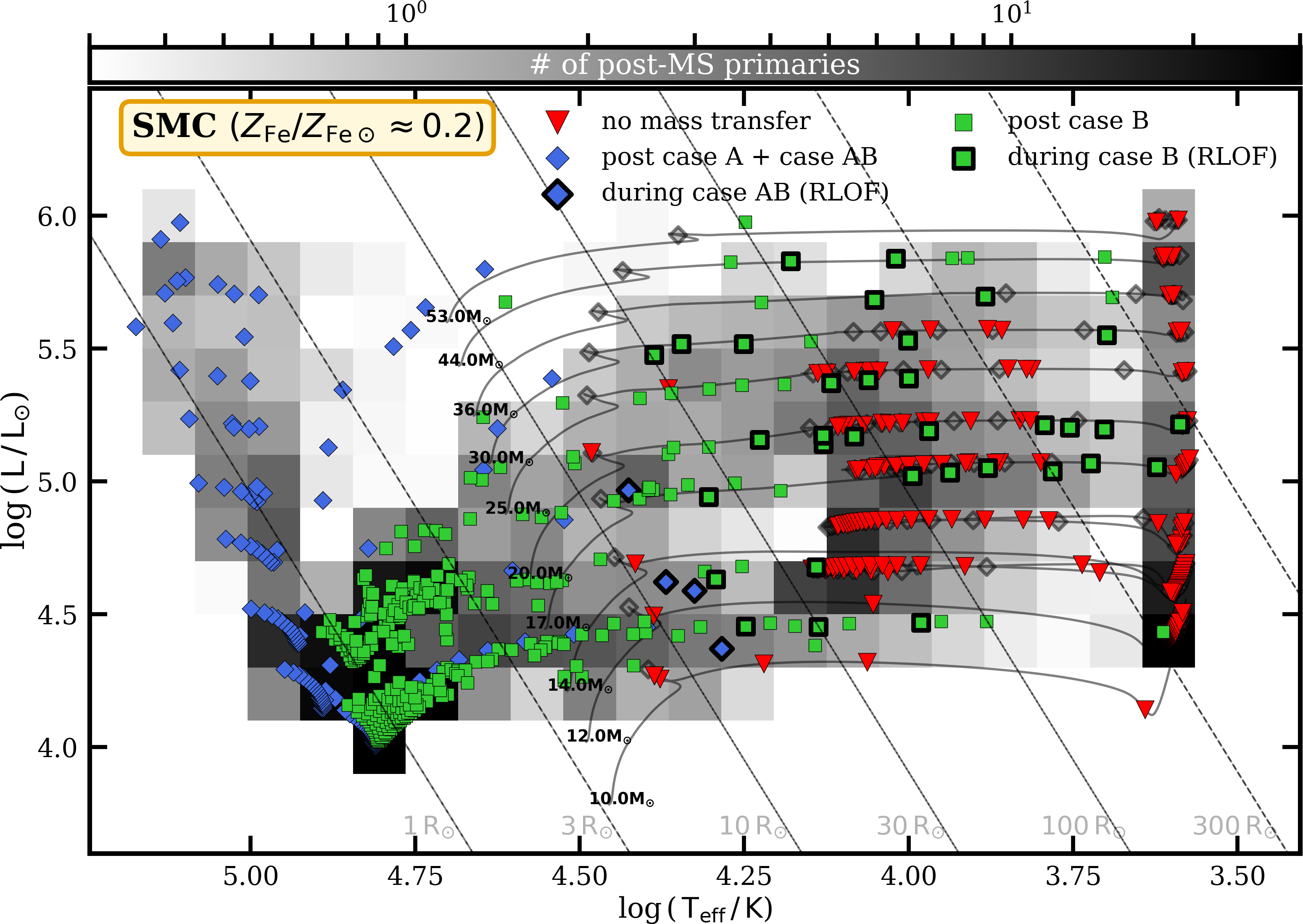}
\caption{Same as Fig.~\ref{fig:disc_HRD_hist2d} but for the SMC composition.}
\label{fig:disc_HRD_hist2d_SMC}
\end{figure*}
  
\begin{figure*}
\includegraphics[width=\textwidth]{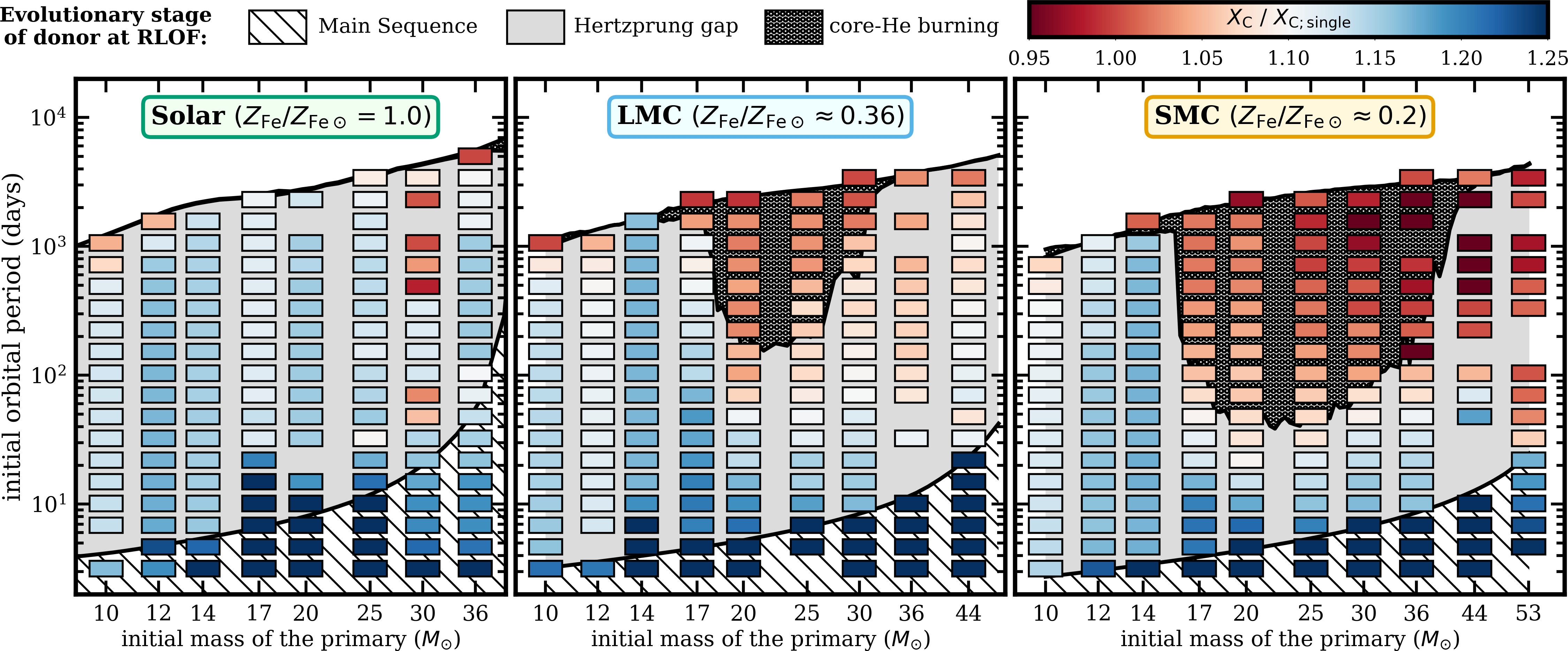}
\caption{Same as Fig.~\ref{fig:square_COcore_mass} but comparing the central carbon mass fraction $X_{\rm C}$ between the single and binary models, rather than the CO core mass.}
\label{fig:square_Xc}
\end{figure*}

 \end{appendix}

\end{document}